\documentclass[twocolumn,showpacs,preprintnumbers,amsmath,amssymb,prl,superscriptaddress]{revtex4}
\usepackage{graphicx}
\usepackage{dcolumn}
\usepackage[tight]{subfigure}
\usepackage{amsmath}
\usepackage{verbatim}
\usepackage[usenames,dvipsnames]{color}
\usepackage{bm} 
\usepackage{bbm}
\usepackage{floatrow}
\usepackage{bibunits}

\newcommand{\Figure}[2]{
  \begin{figure}[t]
    \includegraphics[width=1.0\linewidth]{#1}
    \caption{#2}
  \end{figure}
}
\newcommand{\Bigfigure}[2]{
  \begin{figure*}[ht]
    \includegraphics[width=1.0\linewidth]{#1}
    \caption{#2}
  \end{figure*}
}

\newcommand{\Eq}[1]{Eq.~(\ref{#1})}

\newcommand{\new}[1]{\textcolor{blue}{#1}}
\newcommand{\newer}[1]{\textcolor{OliveGreen}{#1}}
\newcommand{\cut}[1]{\textcolor{red}{cut: #1}}
\newcommand{\todo}[1]{ \textbf{\textcolor{Bittersweet}{TODO: #1}} }
\newcommand{\hide}[1]{ \textbf{\textcolor{Gray}{#1}} }
\newcommand{\tmtextit}[1]{ \textit{#1}}
\newcommand{\nocomma}{}
\newcommand{\tmem}[1]{{\em #1\/}}
\newcommand{\tmop}[1]{\ensuremath{\operatorname{#1}}}

\newcommand{\nobracket}{}

\renewcommand{\new}[1]{{#1}}
\renewcommand{\newer}[1]{{#1}}
\renewcommand{\cut}[1]{{}}
\renewcommand{\todo}[1]{}
\renewcommand{\hide}[1]{}

\begin{document}
\begin{bibunit}[apsrev]

\title{Distinguishing Majorana bound states \\ from localized Andreev bound states by interferometry}

\author{Michael Hell}
\affiliation{Center for Quantum Devices and Station Q Copenhagen, Niels Bohr Institute, University of Copenhagen, DK-2100 Copenhagen, Denmark}
\affiliation{Division of Solid State Physics and NanoLund, Lund University, Box.~118, S-22100, Lund, Sweden}
\author{Karsten Flensberg}
\affiliation{Center for Quantum Devices and Station Q Copenhagen, Niels Bohr Institute, University of Copenhagen, DK-2100 Copenhagen, Denmark}
\author{Martin Leijnse}
\affiliation{Center for Quantum Devices and Station Q Copenhagen, Niels Bohr Institute, University of Copenhagen, DK-2100 Copenhagen, Denmark}
\affiliation{Division of Solid State Physics and NanoLund, Lund University, Box.~118, S-22100, Lund, Sweden}

\date{\today}

\begin{abstract}
  Experimental evidence for Majorana bound states (MBSs) is so far mainly based on the robustness of a zero-bias conductance peak. However,
  similar features can also arise due to Andreev bound states (ABSs) localized at
  the end of an island. We show that these two scenarios can be distinguished
  by an interferometry experiment based on embedding a Coulomb-blockaded
  island into an Aharonov-Bohm ring.
  For two ABSs, when the
  ground state is nearly degenerate, cotunneling can change the state of the island and interference is suppressed.
  By contrast, for two MBSs the ground state is nondegenerate and cotunneling has to preserve the island state, which leads to $h / e$-periodic conductance oscillations with
  magnetic flux. Such interference setups can be realized with semiconducting
  nanowires or two-dimensional electron gases with proximity-induced
  superconductivity and may also be a useful spectroscopic tool for
  parity-flip mechanisms. 
\end{abstract}

 \maketitle


  \Figure{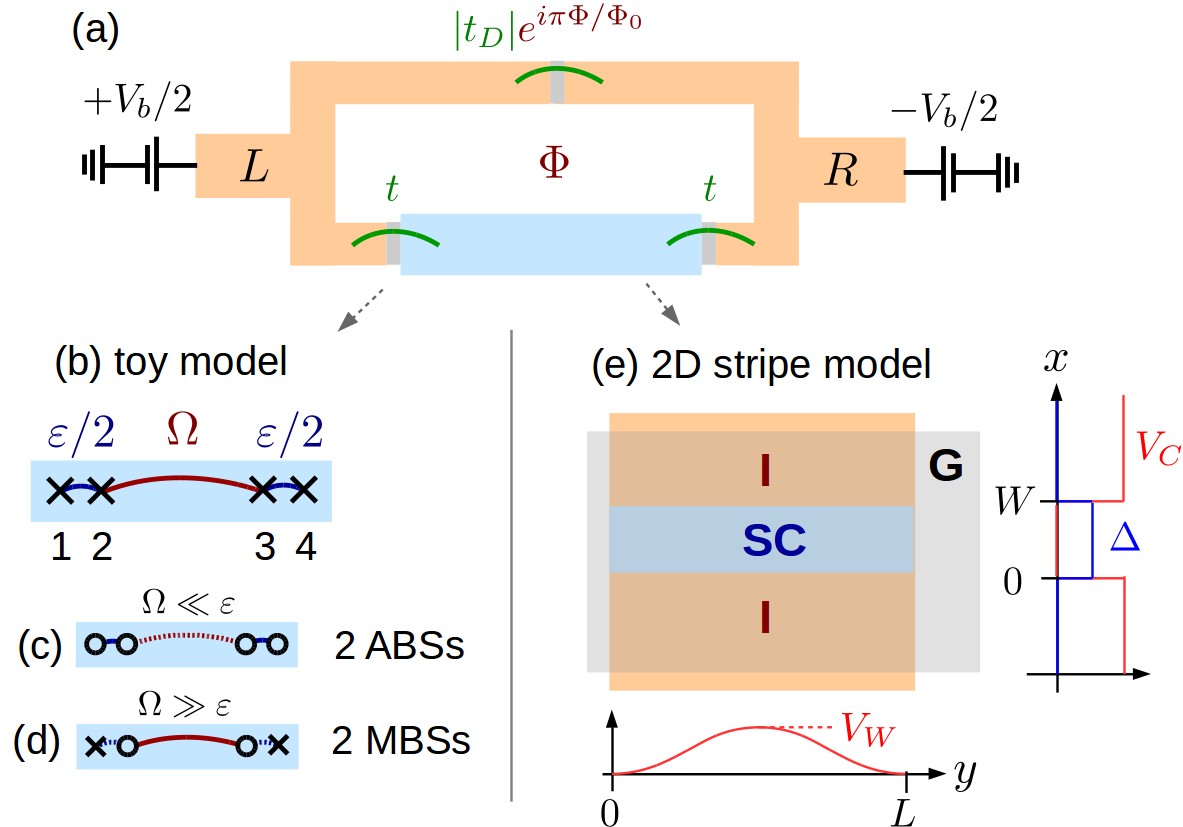}{Sketch
  of interferometer model. (a) Two normal conducting leads (orange, labeled
  $L$ and $R$) are connected via a superconducting island (blue) and a
  reference arm. (b) Toy model for the island consisting of four Majoranas,
  tunable from (c) two terminal ABSs to (d) two terminal MBSs. (e) 2D model for
  a Majorana stripe as seen from top: A stripe of superconductor (SC, blue) is
  placed on top of a semiconductor (orange) and induces a superconducting gap (blue in
  side graph). A gate on top (G, gray) induces a transverse confinement
  potential $V_C$ (red in side graph). Increasing a potential barrier $V_W$
  (red in bottom graph) along the stripe tunes the stripe from case
  (d) to (c). \new{Majoranas that are weakly (strongly) coupled to others are depicted by \newer{crosses} (circles with a conneting line).} \label{fig:model}}

Andreev bound states (ABSs) are coupled particle--hole excitations of
superconductors bound to impurities {\cite{Yu65,Shiba68,Rusinov69}}, to their
surface {\cite{Lofwander01}}, or in junctions {\cite{Kulik69}} with an energy
in the superconducting gap. Since an ABS is a fermionic excitation, its field
operator $f = \gamma_1 + i \gamma_2$ can be decomposed into a pair of
Majorana operators $\gamma_1 = \gamma_1^{\dag}, \gamma_2 = \gamma_2^{\dag}$. While the corresponding wave functions
overlap in space in most cases, they can also be spatially separated for topological superconductors with
triplet pairing
{\cite{1DwiresKitaev,AliceaReview,FlensbergReview,BeenakkerReview,TewariReview}}.
This pins the energy of these Majorana bound states (MBSs) robustly to the
middle of the superconducting gap and renders their non-Abelian exchange
statistics accessible through braiding
{\cite{AliceaBraiding,ClarkeBraiding,HalperinBraiding,BeenakkerBraiding,Aasen16,Hell16,SauBraiding,BraidingWithoutTransport,BondersonBraiding,Vijay16,Karzig17}}.
Both properties may be useful for quantum computation
{\cite{BravyiKitaev,BravyiKitaev2,Freedman03}}.

Topological superconductors may be realized in semiconductors with strong spin-orbit
coupling, proximity-induced superconductivity, and magnetic fields
{\cite{1DwiresOreg,1DwiresLutchyn}}. Evidence for MBSs in these systems is based on a robust zero-bias conductance peak {\cite{mourik12,das12,finck12,Rokhinson,Deng12,Churchill13,Albrecht16,Deng16,Suominen17,Nichele17}} as predicted by theory {\cite{ZeroBiasAnomaly1,ZeroBiasAnomaly3,ZeroBiasAnomaly4,ZeroBiasAnomaly5,ZeroBiasAnomaly6}}.
However, such a peak can also be caused by disorder
{\cite{MajoranaImposter2}}, multi-band effects {\cite{MajoranaImposter3}},
weak antilocalization {\cite{MajoranaImposter4}}, the Kondo effect
{\cite{Goldhaber98}} and, in particular, ABSs
{\cite{MajoranaImposter1,MajoranaImposter5}}. To rule out disorder effects,
intensive efforts have been made to fabricate cleaner devices
{\cite{Marcus15_NatureMat_14_400,Marcus15_NatureNano_10_232,Zhang17,Gazibegovic17,Gul17}}.


Distinguishing MBSs from ABSs is one of the most
urgent goals in Majorana research.
What we refer to here as ABSs are modes with a large Majorana overlap.
If ABSs are extended along the island, they may be discriminated from MBSs by probing a finite conductance in the middle of the island or by a strong response to a gate affecting the middle region. However, one cannot discriminte ABSs from MBSs in this way if there are two terminal ABSs, i.e., one ABS localized at each end of
an island [Fig. \ref{fig:model}(c)].
While the general expectation is that ABSs do not show a similar robustness against parameter
variations as MBSs, ABSs can stick close to zero energy under special conditions when the longitudinal confinement potential is smooth {\cite{Kells12,Liu17}}. This situation has to be contrasted with the desired situation of two MBSs [Fig. \ref{fig:model}(d)] when the potential is rather uniform and rises sharply at the end
of the island.

In this \new{Rapid Communication}, we show how to distinguish the case of two terminal ABSs close
to zero energy from the case of two terminal MBSs by embedding a
Coulomb-blockaded island into an interferometric setup [Fig.
\ref{fig:model}(a)]. Interferometers have been proposed earlier to detect MBSs in grounded
{\cite{Benjamin10,EnMing14,Sun14,Ueda14,Ueda14b,Tripathi16,Dahan17}} and floating
{\cite{Fu10,Yamakage14,Sau15Nonlocal,Rubbert16,Chiu17}} devices and also to
distinguish MBSs from ABSs {\cite{Tripathi16,Sau15Nonlocal}}.
The advantages of our proposal are that it (i) relies on a standard charge current measurement, (ii) successfully distinguishes between MBSs and ABSs also when the MBSs are not fully localized, and (iii) can straightforwardly be implemented using current fabrication capabilities. 

We focus on the case when the charging energy $E_C$ is
the dominant energy scale (besides the superconducting gap $\Delta$) as in
Majorana box qubits {\cite{Plugge17,Plugge16}}. This allows us to study the transport in the cotunneling regime when the total charge on the
island is fixed. This also fixes the total fermion parity of the ground state,
which can be (almost) two-fold degenerate in the case of two ABSs, while it is
nondegenerate for two MBSs. Thus, cotunneling processes cannot change the state
of the island for two MBSs and allow for a large interference
contrast. This is different from the limiting case of two localized ABSs, in which the parity of
both ABSs can be flipped \cite{Sau15Nonlocal}. This conserves the total fermion parity and reduces the
interference contrast strongly.
We show that this mechanism,
captured by a toy model [Fig. \ref{fig:model}(b)], also holds when using a
microscopic 2D model of the island [Fig.
\ref{fig:model}(e)] tuning between the two limits.


{\tmem{Toy model.}} Let us consider \new{an island} that hosts four
Majoranas 1, $\ldots$ 4, two localized at each end [Fig.
\ref{fig:model}(b)]. The Hamiltonian reads
\begin{eqnarray}
  H_I & = & i \varepsilon ( \gamma_1 \gamma_2 + \gamma_3 \gamma_4) - i \Omega
  \gamma_2 \gamma_3 + E_{C, n},  \label{eq:ham}
\end{eqnarray}
where we included the charging energy of the island $E_{C, n} = E_C ( n -
n_g)^2$. Here, $n$ is the number operator for the electrons on the island and
$n_g$ describes the gating. The above toy model interpolates between the
situation of two terminal ABSs and two terminal MBSs: When $\Omega \ll
\varepsilon$, two ABS are at energy $\approx \varepsilon$ [Fig.
\ref{fig:model}(c)]. As they are formed predominantly by the Majorana operators $(
\gamma_1, \gamma_2)$ and $( \gamma_3, \gamma_4)$, we will denote them by $\langle 12 \rangle$
and $\langle 34 \rangle$, respectively. By contrast, when $\Omega \gg \varepsilon$, there
are two terminal MBSs [Fig. \ref{fig:model}(d)]. The corresponding Majorana operators ($\gamma_1$, $\gamma_4$) form a mode  $\langle 14 \rangle$ with a small
energy $\approx \text{ } \varepsilon^2 / 2 \Omega$. In addition, the pair of Majorana operators ($\gamma_2$, $\gamma_3$)
forms a mode  $\langle 23 \rangle$ at higher energy $\approx 2 \Omega$.

\tmtextit{Interferometer model.} The interferometer is enclosed between two
nonsuperconducting leads described by $H_0 = \sum_{r k \sigma} (
\varepsilon_{r k} - \mu_r) c^{\dag}_{r k \sigma} c_{r k \sigma}$, where $c_{r
k \sigma}$ denotes the annihilator for electrons in lead $r = L, R$ in mode
$k$ with spin $\sigma = \uparrow, \downarrow$. The leads are held at a common
temperature $T$ and are voltage-biased symmetrically: $\mu_L = - \mu_R = V_b /
2$ (We set $e = \hbar = c = k_B = 1$).

The tunnel Hamiltonian reads
\begin{eqnarray}
  H_T & = & \sum_{r k \sigma m=1,2} c_{r k \sigma} e^{i \varphi / 2} t_{r \sigma
  m} ( \delta_{r L} \gamma_m + \delta_{r R} \gamma_{5 - m}) \nonumber\\
  &  & + \sum_{k k' \sigma \sigma'} t_{D, \sigma \sigma'} c^{\dag}_{L k
  \sigma} c_{R k' \sigma'} + \text{H.c.,}  \label{eq:ht}
\end{eqnarray}
where $\varphi$ denotes the superconducting phase on the island and $m=1,2$ enumerates the Majorana operators. In our toy
model, we assume that lead $r$ couples only to the two nearest MBSs [first line
of Eq. (\ref{eq:ht})] with energy-independent tunnel matrix elements $t_{r
\sigma m}$. \new{For simplicity, we} assume the island to be left-right symmetric, \new{so that} they obey
the relation $t_{L \sigma m} = ( - 1)^m \sigma t_{R \bar{\sigma}  \bar{m}} =
t_{\sigma m}$ {\cite{Hell17csup}}. By rotating the spin basis in
the leads, one can parametrize the tunnel matrix elements conveniently as
$t_{\uparrow 1} = t \cos ( \lambda)$, $t_{\downarrow 1} = 0$, $t_{\uparrow 2}
= t \sin ( \lambda) \cos ( \beta) e^{i \delta}$, and $t_{\downarrow 2} = t
\sin ( \lambda) \sin ( \beta) e^{i \delta}$ {\cite{Hell17csup}}. The parameter
$t$, \new{together with the spin- and energy-independent density of states $\nu$ of the leads} sets the overall tunnel rate \ $\Gamma = 2 \pi \nu | t |^2$ between the
leads and the island, $\lambda$ characterizes the relative coupling strength
of the two Majoranas to the leads, $\delta$ is a relative phase shift, and
$\beta$ is the canting of the different spin directions the two Majoranas couple
to.

In our model, a featureless reference arm connects the two leads
[second term in Eq. (\ref{eq:ht}).
The phase of the direct tunnel
amplitude $t_{D,\sigma \sigma'} = | t_D | ( \delta_{\sigma \sigma'} + \tau_{\tmop{sf}}
\delta_{\sigma \bar{\sigma}'}) e^{i \pi \Phi / \Phi_0}$ is controlled by the
magnetic flux $\Phi$ threaded through the loop ($\Phi_0 = e / 2 h$).
\new{We neglect here decoherence in the reference arm, which is motivated by the experimental observation of phase-coherent transport up to several $\mu$m in InAs \cite{Yang02,Gazibegovic17} and InGaAs \cite{Ren13} interferometers.}
Note that
if $\lambda = 0$ or $\beta = 0$, the island couples only to electrons with
spin $\uparrow$ ($\downarrow$) in the left (right) lead. In the special case when the tunneling in
the reference arm is spin-conserving ($\tau_{\tmop{sf}} = 0$), no interference
can appear because one can tell from the spin of the outgoing electron which
path has been taken {\cite{Akera93}}. In practice, the island is of course not
perfectly symmetric and spin-orbit coupling rotates the spin of electrons
traveling through the reference arm, resulting in a nonzero interference. For
simplicity, we set $\tau_{\tmop{sf}} = 1$, which limits the interference contrast to 1/2 when $\beta=0$ [\Eq{eq:hmax}].

\tmtextit{Transport calculations.} Our goal is to understand the behavior of
the maximal interference contrast:
\begin{eqnarray}
  \text{MIC} & : = & \underset{\Phi, | t_D |}{\max}  \left| \frac{I ( \Phi) - I (
  \Phi + \Phi_0)}{I ( \Phi) + I ( \Phi + \Phi_0)}\right| .  \label{eq:ic}
\end{eqnarray}
Here, $I ( \Phi)$ is the stationary current through the interferometer. Note that the maximal or minimal current may not necessarily flow for $\Phi = 0, \pi$.
Since
interference requires coherent transport through the island, we constrain our
calculations to the cotunneling regime. We \new{set up a master equation  \cite{Hell17csup} and} consider the specific situation
when only one particular charge state $n = n_0$ of the island is occupied and
cotunneling predominantly involves only the adjacent charge state $n_0 + 1$
($\Gamma, T, V_b \ll U = E_{C, n_0 + 1} - E_{C, n_0} \ll E_{C, n_0} - E_{C,
  n_0 - 1}$). \new{Without loss of generality, we assume $n_0$ to be even.} While our toy model neglects cotunneling through the
quasiparticle continuum, quasiparticle states are included partially later on in the 2D island model.
The cotunneling rates are computed with the T-matrix approach \new{ including terms of $O ( t^2, t_D)$ into the T-matrix {\cite{Hell17csup}}. We neglect all other contributions, including those} leading to the Kondo
effect ($\Gamma, T_K \ll T$) \new{and} Cooper-pair cotunneling forming a virtual
intermediate Cooper pair ($\Gamma \ll U, \Delta$).






{\tmem{Interference contrast for toy model.}} To contrast the cases of two
MBSs and two ABSs, we
first study the parameter dependence of the MIC for the toy model (\ref{eq:ham}). When $\beta = \delta = 0$ and $V_b \ll E = \sqrt{\varepsilon^2+\Omega^2} \ll U$, the
MIC reads
\begin{eqnarray}
  \text{MIC} & = & \frac{\tanh ( E / T)}{2 \sqrt{1 + \left( \frac{E /
  \Omega}{\cos^2 ( 2 \lambda)} - 1 \right) \frac{2 E / T}{\sinh ( 2 E / T)}}}
  .  \label{eq:hmax}
\end{eqnarray}
Details including an expression for general bias voltage are given in
{\cite{Hell17csup}}. 
We see that the MIC tends to its maximal value 
when $\Omega / \varepsilon \gg 1$ (two MBSs), while it tends to zero when
$\Omega / \varepsilon \ll 1$ (two ABSs) [Fig. \ref{fig:toy-model-par-dep}(a)].
This implies that the case of two MBSs and two ABSs can be distinguished by the
maximally achievable MIC. In the next two paragraphs, we explain the different behavior of the two cases when only Majoranas
1 and 4 are connected to the leads ($\lambda = 0$).

  \Figure{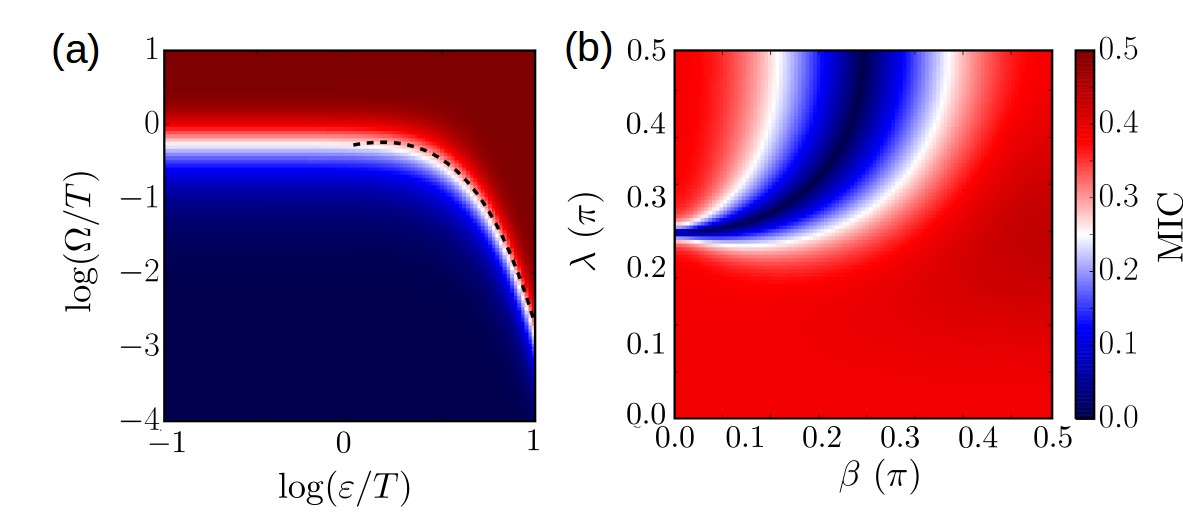}{Parameter
  dependence of the MIC for the toy model. (a) Dependence on
  Majorana coupling energies for $\lambda = \beta = 0$. The black-dashed line
  is given by $\Omega = \sqrt{2 \varepsilon^3 / T} e^{- \varepsilon / T}$ and
  marks the crossover between transport dominated by parity-conserving and parity-flipping
  cotunneling. (b) Dependence on tunnel matrix elements for $\Omega /
  T = 1$, $\varepsilon / T = 0.1$. In both cases, we used $V_b = 0.01 T$,
  $U = 100 T$, and $\delta = 0$.\label{fig:toy-model-par-dep}}

When $\Omega / \varepsilon \gg 1$ and $\Omega \gg V_b, T$, the island resides
mostly in its ground state, in which the parities of the modes  $\langle 23 \rangle$ and  $\langle 14 \rangle$ are even. Transport is
predominantly carried by parity-conserving cotunneling processes: An electron incoming
from one lead flips the parity of mode  $\langle 14 \rangle$ and the outgoing electron
flips it back. Such electrons interfere with electrons tunneling through the
reference arm and lead to a large MIC [Fig. \ref{fig:toy-model-par-dep}(a)].
The MIC is suppressed when voltage bias or temperature exceed the
inelastic cotunneling threshold, i.e., when $\text{min} ( V_b, T) > E$ [Fig.
\ref{fig:toy-model-par-dep}(a)]. In this case cotunneling processes
can flip the parity of the modes  $\langle 14 \rangle$ and $\langle 23 \rangle$ and bring the island from its ground state to
the excited state. We will refer to this as parity-flipping processes (referring to the individual modes) even though the total fermion parity of the island is of course preserved.
The occupation probability of the ground and excited state
tend to 1/2 when $\text{min} ( V_b, T) \gg E$. Importantly, the flux dependence of the
cotunneling rates differs by $\pi$ depending on the initial parity of mode  $\langle 23 \rangle$ in the cotunneling process {\cite{Fu10}}. Hence, interference is still possible
in each cotunneling event but the MIC becomes suppressed due to averaging over
both possible initial states.

When $\Omega / \varepsilon \ll 1$, the MIC can be suppressed even if $E \gg \text{max} ( V_b, T)$.
The reason is that parity-conserving cotunneling is
strictly forbidden in the limit $\Omega = 0$: The left lead couples only to
mode $\langle 12 \rangle$, while the right lead only couples to mode  $\langle 34 \rangle$. A cotunneling
process transferring an electron from one to another must therefore flip the
parities of both modes and thus results in the final state being different from the initial state. Hence, there is no interference. When $\varepsilon > T$, the crossover from transport dominated by parity-conserving to parity-flipping processes happens when $\Omega > \sqrt{2 \varepsilon^3 / T} e^{- \varepsilon / T}$
($V_b \ll T$) [Fig.
  \ref{fig:toy-model-par-dep}(a)]. In experiments, this crossover may be
influenced by other processes that can flip the parities of the ABSs, such as
quasiparticle poisoning from the continuum {\cite{Higginbotham15}}, or
Cooper-pair splitting due to photons {\cite{Bretheau13}} or phonons
{\cite{Patel16}}. \newer{If the current is averaged over a time shorter than the time between two parity flips, interference remains detectable and the parity of  mode $\langle 23 \rangle$ can be read out \cite{Plugge17}. The MIC may thus also be utilized to measure such rates. The results we show here have to be understood as long-time averages of many parity flips instead.}

  \Bigfigure{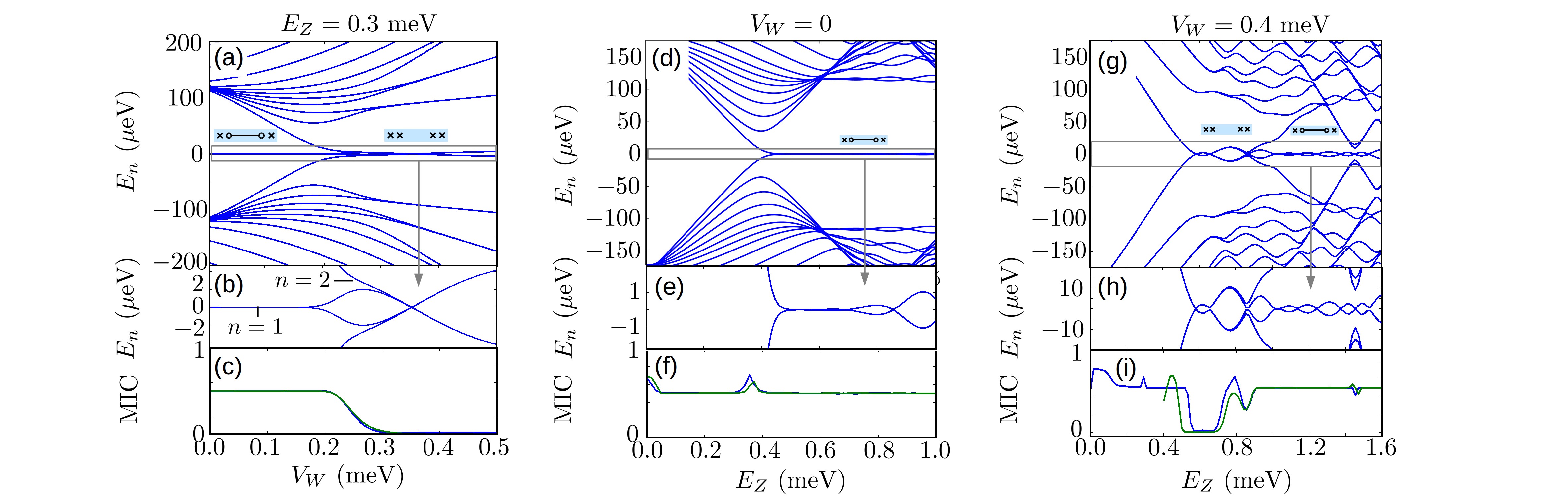}{Suppression of the interference contrast for the transition from two MBSs to two ABSs.
   We show the energy spectrum
  of the stripe Hamiltonian (\ref{eq:ham}) (upper panels) with close-ups around zero
  energy (middle panels) alongside the MIC (lower panels). We compute the MIC both for the 2D stripe model (blue) and the toy model
  with paraxmeters extracted from the 2D stripe model (green) {\cite{Hell17csup}}. The values of $E_Z$ and $V_W$ are specified in the panels,
  the lattice constant is $a = 10~\text{nm}$, $\Delta = 180~\mu \text{eV}$, $E_{\tmop{SO}} = m^{\ast}
  \alpha^2 / 2 = 116.5 ~\mu \text{eV}$, $\mu = 0$, $V_C = 1~
  \tmop{meV}$, $W = 200~\text{nm}$, $L = 2~\mu\text{m}$, $m^{\ast} = 0.023~m_e$, $U
  = 50~\mu \text{eV}$, \ $T = 1.6~\mu \text{eV} ~( \approx \text{20 mK})$,
  and $V_b = 1~\mu \text{eV}$. In (i), the toy model breaks down for
  $E_Z \lesssim 0.4~$meV (see text).  \label{fig:v0-dep}}

The qualitative parameter dependence of the MIC remains in most cases
unchanged if one considers the general case of $\lambda \neq 0, \beta \neq 0,
\delta \neq 0$. From numerical calculations, we find only a weak
dependence of the MIC on $\delta$ except for special points \cite{Hell17csup}. We find, however, a suppression
of the MIC under the condition $\sqrt{( \lambda - \pi / 2)^2 + \beta^2}
\approx \pi / 4$ [Fig. \ref{fig:toy-model-par-dep}(b) and Eq.
(\ref{eq:hmax})]. Here, the parity-conserving cotunneling rates vanish because of
destructive interference of processes involving only the island (not
the reference arm).  We finally note that the case $\Omega = 0$ with $\varepsilon=0$ or $\lambda=0$ is a pathological case of our model \cite{Hell17csup}.


{\tmem{2D model for Majorana stripe}}. To see whether the simple toy model
discussed so far indeed captures the main physics to contrast the cases of two MBSs and two ABSs, we next turn to a more sophisticated model for the island.
Following {\cite{Hell17a}}, we consider a Majorana stripe of width $W$ and
length $L$ defined in a two-dimensional electron gas [Fig.
\ref{fig:model}(e)]. The electron gas is modeled by a single electron band
with effective mass $m^{\ast}$ at chemical potential $\mu$ as described by the
following Bogoliubov-de Gennes Hamiltonian:
\begin{eqnarray}
  \mathcal{H}_{\text{BdG}} & = & \left( - \frac{\partial_x^2 + \partial_y^2}{2 m^{\ast}} + V_C ( x) +
  V_W ( y) - \mu \right) \tau_z  \label{eq:ham}\\
  &  & - i \alpha ( \sigma_x \partial_y - \sigma_y \partial_x) \tau_z + E_Z
  \sigma_y / 2 + \Delta ( x) \tau_x. \nonumber
\end{eqnarray}

In the second line, we added the Rashba spin-orbit coupling (with velocity
$\alpha$), the Zeeman energy ($E_Z$) due to a magnetic field, and the induced
superconducting gap. The latter is nonzero where the electron gas is covered
by the superconductor: $\Delta ( x) = \Delta \Theta ( W / 2 - | x |)$. The
Hamiltonian acts on the four-component spinor $[u_{\uparrow} ( x, y), u_{\downarrow} ( x, y), v_{\downarrow}
( x, y), - v_{\uparrow} ( x, y)]^T$ containing the electron $( u)$ and
hole $( v)$ components for spin $\sigma = \uparrow, \downarrow$. The Pauli
matrices $\tau_i$ and $\sigma_i$ ($i = x, y, z$) act on particle-hole and spin
space, respectively. Gates are used to confine the states in the transverse
direction, $V_C ( x) = V_C \Theta ( | x | - W / 2)$ ($V_C \gg \mu, \Delta,
E_Z, E_{\tmop{SO}} = m \alpha^2 / 2$). Equation (\ref{eq:ham}) also models a
nanowire if the transverse confinement in one direction is much stronger than
in the other (e.g. due to gating)
{\cite{Lutchyn11Multiband,Multichannel1,Disorder7}}. 

{\tmem{Tuning from two MBSs to two ABSs.}} \ Accounting for an additional
potential profile along the stripe, $V_W ( y)  = V_W [ 1 + \cos ( 2 \pi y / L)]$, 
we can tune from the case of two MBSs to two ABSs by increasing $V_W$.
Computing the energy
spectrum of the island [Fig. \ref{fig:v0-dep}(a)], we find for $V_W = 0$ only one
mode ($n = 1$) close to zero energy. This mode is formed by two slightly
overlapping MBSs at opposite ends of the stripe. When increasing $V_W$, the second mode ($n=2$) comes close to and sticks to zero energy [Fig. \ref{fig:v0-dep}(b)]. When $V_W$ is large, the two modes correspond to two ABSs localized at the ends of the wire \cite{Hell17csup}. The MIC is reduced when $V_W$ is increased as the system
evolves from two MBSs to two ABSs [Fig. \ref{fig:v0-dep}(c)].

To compute the MIC using the 2D model, we
include the 8 lowest modes into our master equation approach. We further
extracted the parameters for the toy model from the energies (yielding
$\Omega$ and $\varepsilon$) and wave functions (yielding the tunnel matrix
elements) of the two lowest modes obtained for the 2D model. In this
extraction procedure {\cite{Hell17csup}}, we neglect the coupling of the Majoranas
on the left (right) to the right (left) lead. We find that the toy model
reproduces the MIC rather accurately.


We finally discuss the magnetic-field dependence of the MIC [Figs.
\ref{fig:v0-dep}(d)--(i)]. Similar to the case of two MBSs, the energies of the two ABSs
oscillate around zero energy as a function of magnetic
field [compare Figs. \ref{fig:v0-dep}(e) and (h)]. For $V_W = 0$,
we see that the MIC also stays large in the nontopological regime for small values of $E_Z$ [Fig. \ref{fig:v0-dep}(f)]. The
reason is that parity-flipping processes are energetically forbidden as long as $2 ( E_1 + E_2) \ll V_b, T$.
However, when $E_Z$ is small, the
Coulomb peaks are not 1$e$ periodic \cite{Albrecht16}, which is a way to distinguish the
nontopological from the topological regime in this case.

For the case of two ABSs ($V_W=0.4~$meV), we find that
the MIC is suppressed when there are two modes close to zero energy [Fig.
\ref{fig:v0-dep}(i)]. The MIC is restored again when at least one of the modes
has an energy $\gg V_b, T$. This happens for small magnetic fields [$E_Z < 0.5 \text{ meV}$ in Fig.
\ref{fig:v0-dep}(i)] when the ABSs are at large energies or when the case of two MBSs is restored [$E_Z > 0.9 \text{
    meV}$ in Fig. \ref{fig:v0-dep}(i)]. Again, for small $E_Z$, the Coulomb peaks are not 1$e$ periodic, which rules out the presence of MBSs.
We note that the toy model breaks down in this regime because
$\Omega / \varepsilon$ becomes very small (leading to nearly zero current through the island). This does not happen for the full 2D model where all tunnel couplings are accounted for.

{\tmem{Conclusion.}} A zero-bias conductance peak in transport spectroscopy of
superconducting islands can arise due to MBSs as well as ABSs. While
extended ABSs may be probed by a contact in the middle of a superconducting
stripe, terminal ABSs cannot. We have shown that terminal ABSs can instead be
distinguished from two terminal MBSs by an interference experiment. Such
experiments may also be useful to probe quasiparticle-poisoning rates for
nonisolated islands.
Finally, the idea of our approach may be of interest for initial testing of the presence of MBS in Majorana-qubit devices \cite{Plugge17,Karzig17}, in which interferometers are integrated as a means of readout.

\begin{acknowledgments}
{We acknowledge stimulating discussions with M. Deng, A. Fornieri, L. Glazman, C. M. Marcus, F. Nichele, E. O'Farrell, S. Plugge, A. Stern, A. Whiticar, and support from the Crafoord Foundation (M. L. and M. H.), the Swedish Research Council (M. L.), The Danish National Research Foundation, and from the Microsoft Station Q Program. }
\end{acknowledgments}

\putbib[cite]
\end{bibunit}


\clearpage
\begin{bibunit}[apsrev]
  \setcounter{equation}{0}
  \setcounter{figure}{0}
  \setcounter{section}{0}
  \setcounter{affil}{0}

\renewcommand{\thefigure}{S\arabic{figure}}
\renewcommand{\theequation}{S\arabic{equation}}

\title{Distinguishing Majorana bound states from localized Andreev bound states by interferometry: Supplemental Material}
\author{Michael Hell}
\affiliation{Center for Quantum Devices and Station Q Copenhagen, Niels Bohr Institute, University of Copenhagen, DK-2100 Copenhagen, Denmark}
\affiliation{Division of Solid State Physics and NanoLund, Lund University, Box.~118, S-22100, Lund, Sweden}
\author{Karsten Flensberg}
\affiliation{Center for Quantum Devices and Station Q Copenhagen, Niels Bohr Institute, University of Copenhagen, DK-2100 Copenhagen, Denmark}
\author{Martin Leijnse}
\affiliation{Center for Quantum Devices and Station Q Copenhagen, Niels Bohr Institute, University of Copenhagen, DK-2100 Copenhagen, Denmark}
\affiliation{Division of Solid State Physics and NanoLund, Lund University, Box.~118, S-22100, Lund, Sweden}
\date{\today}

 \maketitle

\section{Details of the models}\label{sec:model}

\subsection{2D island model: Eigenmodes and energies}\label{sec:2dmodel}

Since we assume that the charging energy $E_C$ is the dominant energy scale, we
work in the basis of many-body eigenstates denoted by $| n, \pmb{\eta}
\rangle$. Here, $n$ is the number of electrons on the island and
$\pmb{\eta} = ( \eta_1, \eta_2, \ldots)$ contains the occupations
$\eta_l$ of the eigenmodes $l$ of the island. These modes are found by solving
for the eigenstates of the BdG Hamiltonian (5): $\mathcal{H_{\tmop{BdG}}}
\pmb{\chi}_l = \varepsilon_l \pmb{\chi}_l$. The annihilation operator for eigenmode $l$ is given
by
\begin{eqnarray}
  \beta_l & = & \int dx \int dy \ \pmb{\chi}_l^{\dag} ( x, y) \cdot
  \pmb{\psi} ( x, y),  \label{eq:betafield}
\end{eqnarray}
where
$\pmb{\psi} ( x, y) = [ \psi_{\uparrow} ( x, y), \psi_{\downarrow} ( x, y), \psi^{\dag}_{\downarrow}
( x, y), - \psi^{\dag}_{\uparrow} ( x, y)]^T$ contains the electron field operators for spin $\sigma = \uparrow, \downarrow$ at position $(x,y)$.
The energies of the many-body eigenstates read
\begin{eqnarray}
  \varepsilon_{n, \pmb{\eta}} & = & E_C ( n - n_g)^2 + \sum_l \eta_l
  \varepsilon_l . 
\end{eqnarray}
For a nonsuperconducting island, the occupations have to satisfy $\sum_l
\eta_l = n$. This condition is lifted in the superconducting case, where they
only have to obey the constraint $( - 1)^{\sum_l \eta_l} = ( - 1)^n$. Since
the modes $l$ are electron-hole superpositions, the occupations $\eta_l$ thus
only specify the fermion parity $( - 1)^{\eta_l}$ of mode $l$ but not the
charge associated with occupying this mode.

\subsection{Toy model: Eigenmodes and energies}

To derive the left-right symmetry relation of the Majorana tunnel couplings in Sec.
\ref{sec:lrsymtoymodel}, we will need the eigenstates and eigenmodes of the toy
model (1). For this purpose, we first rewrite Eq.~(1) as $H_I = \tfrac{1}{2}
\pmb{\alpha}^{\dag} \cdot \mathcal{H}_I \cdot \pmb{\alpha}$, where
$\pmb{\alpha} = ( \alpha_L, \alpha_R, \alpha_L^{\dag},
\alpha_R^{\dag})^T$ contains the field operators of the modes $\alpha_L = (
\gamma_1 + i \gamma_2) / 2$ and $\alpha_R = ( \gamma_3 + i \gamma_4) / 2$. By
diagonalizing the Hamiltonian matrix
\begin{eqnarray}
  \mathcal{H}_I & = & \left(\begin{array}{cccc}
    \varepsilon & \Omega & 0 & \Omega\\
    \Omega & \varepsilon & - \Omega & 0\\
    0 & - \Omega & - \varepsilon & - \Omega\\
    \Omega & 0 & - \Omega & - \varepsilon
  \end{array}\right),  \label{eq:hi}
\end{eqnarray}
we can express the Hamiltonian as $H_I = \tfrac{1}{2} \pmb{\beta}^{\dag}
\cdot \mathcal{H}'_I \cdot \pmb{\beta}$ with $\mathcal{H}_I' =
\tmop{diag} ( \varepsilon_2, \varepsilon_1, - \varepsilon_1, - \varepsilon_2)$
and eigenenergies
\begin{eqnarray}
  \varepsilon_{1 / 2} & = & \sqrt{\varepsilon^2 + \Omega^2} \mp \Omega . 
  \label{eq:energies}
\end{eqnarray}
The field operators of the eigenmodes are again collected in a 4-component vector $\pmb{\beta} = \mathcal{U \cdot} \pmb{\alpha}$,
where
\begin{eqnarray}
  \mathcal{U} & = & \left(\begin{array}{cccc}
    r_+ & r_+ & -r_- & r_-\\
    r_+ & -r_+ & r_- & r_-\\
    r_- & r_- & r_+ & - r_+\\
    - r_- & r_- & r_+ & r_+
  \end{array}\right)  \label{eq:u}
\end{eqnarray}
and $r_{\pm} = \sqrt{1 \pm 1 / \sqrt{1 + ( \Omega / \varepsilon)^2}} / 2$, which satisfies the relation
$r_+^2 + r_-^2 = 1 / 2$.

\subsection{2D island: Tunnel Hamiltonian}\label{sec:ht2d}

Here, we explain how to extract the tunnel matrix elements from the 4-spinor
components of the solutions of the BdG equation for our transport
calculations.

For the transport calculations for the 2D island model, we start from the
standard bilinear tunnel Hamiltonian
\begin{eqnarray}
  H_T & = & \sum_{r k \sigma} \int dx \int dy \ t_r ( x, y) c_{r k \sigma} e^{i
    \varphi / 2} \psi^{\dag}_{\sigma} ( x, y)   \nonumber \\
  & & \text{+ H.c. } .  \label{eq:htfull}
\end{eqnarray}
We use the same notation as in the main paper and introduced the field operator
$\psi^{\dag}_{\sigma} ( x, y)$ creating an electron on the island at position
$( x, y)$ with spin $\sigma$.

Inverting Eq.~(\ref{eq:betafield}) and using the particle-hole symmetry of
$\mathcal{H_{\text{BdG}}} \nocomma$, $[ \mathcal{P},
\mathcal{H}_{\text{BdG}}]_+ = 0$ with $\mathcal{P} = \sigma_y \tau_y
\mathcal{K}$ ($\mathcal{K}$ denotes the complex conjugation), yields
\begin{eqnarray}
  \psi_{\sigma}^{\dag} ( x, y) & = & \sum_{l~\text{unocc.}} [ u^{\ast}_{l \sigma} ( x,
  y) \beta^{\dag}_l + v_{l \sigma} ( x, y) \beta_l] \label{eq:psirepr}. 
\end{eqnarray}
The sum in \Eq{eq:psirepr} includes every particle-hole conjugated mode pair only once. One is, in principle, free to chose which of the modes at energy $\pm \varepsilon_l$ is used.
The choice we use is that the sum runs only over modes $l$ that are unoccupied at zero temperature. In the topologically trivial regime at zero magnetic field, these are all the modes with positive energy. However, when a mode crosses zero energy (such as a MBS), then the occupation of that state changes at zero temperature. When then use the mode at negative energy. 
 We use this choice because then the correct electron-hole components will be used for the transport calculations when inserting \Eq{eq:psirepr} into the tunnel Hamiltonian (\ref{eq:htfull}):
\begin{eqnarray}
  H_T & = & \sum_{r k \sigma p l~\text{unocc.}} t_{r l \sigma}^p c_{r k \sigma} e^{i
  \varphi / 2} \beta_l^p + \text{H.c.}. \label{eq:htbeta}
\end{eqnarray}
Here, $p = \pm$, $\beta^+_l = \beta^{\dag}_l, \beta^-_l = \beta_l$, where we
assume a point-contact coupling of the leads so that
\begin{eqnarray}
  t_{L / R l \sigma}^+ & = & t_{L / R} ( 0, \mp L / 2) u^{\ast}_{l \sigma} (
  0, \mp L / 2), \\
  t_{L / R l \sigma}^- & = & t_{L / R} ( 0, \mp L / 2) v_{l \sigma} ( 0, \mp L
  / 2).
\end{eqnarray}
In Sec. \ref{sec:lrsym}, we show how to express the tunnel amplitudes to
the right lead in terms of those to the left lead [Eqs. (\ref{eq:usym}) and
(\ref{eq:vsym})] provided the island exhibits a spatial inversion symmetry
along the stripe direction.

For our transport calculations, we will consider only two relevant charge
states $n = 0$ and $n = 1$ of the island. Projecting Eq.~(\ref{eq:htbeta}) on
the many-body basis $| n, \pmb{\eta} \rangle$ introduced in Sec.
\ref{sec:2dmodel}, we obtain
\begin{eqnarray}
  \tilde{H}_T & = & \sum_{r k \sigma \pmb{\eta}_i  \pmb{\eta}_f}
  T_{r \sigma}^{\pmb{\eta}_f  \pmb{\eta}_i} c_{r k \sigma} | 1,
  \pmb{\eta}_f \rangle \langle 0, \pmb{\eta}_i | + \text{H.c.} ,
\end{eqnarray}
with tunnel matrix elements
\begin{eqnarray}
  T_{r \sigma}^{\pmb{\eta}_f  \pmb{\eta}_i} & = & [ t^+_{r l \sigma}
  \delta_{\eta_l 0} + t^-_{r l \sigma} \delta_{\eta_l 1}] \prod_{j \neq l}
  \delta_{\eta_{f, j} \eta_{i, j}} . 
\end{eqnarray}
There are two contributions: The first one ($\sim t^+_{r l \sigma} \sim
u^{\ast}_{l \sigma}$) describes a 'usual' tunneling process, which can also occur
in a nonsuperconducting systems: An incoming electron occupies an empty mode
$l$. The second one $( \sim t^+_{r l \sigma} \sim v_{l \sigma})$ happens only
in superconducting systems and describes the formation of a Cooper pair with
the incoming electron, leaving the mode $l$ empty in the final state.

\subsection{2D model: Spatial inversion symmetry}\label{sec:lrsym}

While Eq.~(\ref{eq:htbeta}) is generally valid for any 2D island model, we use
in our calculations a 2D island Hamiltonian [Eq.~(5)] that obeys a spatial
inversion symmetry: It obeys $[ \mathcal{H}_{\text{BdG}}, \mathcal{V}]_- = 0$
with the unitary operator $\mathcal{V} = \mathcal{I}_y \sigma_y$. Here,
$\mathcal{I}_y$ is the inversion along the stripe direction $( y)$ and
$\sigma_y$ is the Pauli matrix acting on spin. Thus, if $\pmb{\chi}_l$ is
a solution of the BdG equation, $\mathcal{H}_{\text{BdG}} \pmb{\chi}_l =
\varepsilon_l \pmb{\chi}_l$, then $\mathcal{V} \pmb{\chi}_l$ is also
a solution for the BdG equation for the same energy. If the solution is
nondegenerate, this implies $\mathcal{V} \pmb{\chi}_l = \xi_l
\pmb{\chi}_l$ with $\xi_l = \pm 1$ because $\mathcal{V}^2 =
\mathcal{V}^{\dag} \mathcal{V} = \mathbbm{1}$. The unitary transformation
$\mathcal{V}$ relates the wave-function components of an eigenstate on the
left side of the island to that on the right side: Expressing
$\pmb{\chi}_l = ( u_{l \uparrow}, u_{l \downarrow}, v_{l \downarrow}, -
v_{l \uparrow})^T$, we obtain
\begin{eqnarray}
  u_{l \sigma} ( x, y) & = & - i \xi_l \sigma u_{l \bar{\sigma}} ( x, - y), 
  \label{eq:usym}\\
  v_{l \sigma} ( x, y) & = & - i \xi_l \sigma v_{l \bar{\sigma}} ( x, - y) . 
  \label{eq:vsym}
\end{eqnarray}
The tunnel couplings thus satisfy the left-right symmetry
\begin{eqnarray}
  t_{r l \sigma}^p & = & i p \xi_l \sigma t^p_{\bar{r} l \bar{\sigma}} . 
  \label{eq:tsym}
\end{eqnarray}
\subsection{Toy model: Spatial inversion symmetry}\label{sec:lrsymtoymodel}

If the toy model is compatible with the 2D island model, then the tunnel
couplings of the eigenstates of the toy model must also obey Eq.~(\ref{eq:tsym}). We next briefly explain how this translates into conditions
for the Majorana tunnel couplings stated below Eq.~(2) in the main paper. \new{This symmetry of the tunnel couplings is, however, not essential to our findings, it is only convenient to reduce the number of parameters.}

In terms of the localized modes $\alpha_L^p = ( \gamma_1 - i p \gamma_2) / 2$
and $\alpha_R^p = ( \gamma_1 - i p \gamma_2) / 2$, the tunnel Hamiltonian can
be expressed in the same form as Eq.~(\ref{eq:htbeta}):
\begin{eqnarray}
  H_T & = & \sum_{r k} \tilde{t}_{r \sigma}^p c_{r k \sigma} e^{i \varphi / 2}
  \alpha_r^p + \text{H.c.} 
\end{eqnarray}
Here we use the assumption that mode $\alpha_r$ couples only to lead $r$.
Exploiting the transformation $\pmb{\beta} = \mathcal{U} \cdot
\pmb{\alpha}$ with $\mathcal{U}$ given by Eq.~(\ref{eq:u}) and expressing
the field operators according to Eq.~(\ref{eq:betafield}), we find the
relation ($l=1,2$)
\begin{eqnarray}
  t^p_{r l \sigma} & = & ( - 1)^{\delta_{r, R} l} r_+  \tilde{t}_{r \sigma}^p
  + ( - 1)^{\delta_{r, L} ( l - 1)} r_- \tilde{t}_{r \sigma}^{\bar{p} \ast} . 
\end{eqnarray}
Using furthermore Eq.~(\ref{eq:tsym}), we obtain for $\xi_1 + \xi_2 = 0$
\begin{eqnarray}
  \tilde{t}_{r \sigma}^p & = & - i p \sigma \xi \tilde{t}_{\bar{r} 
  \bar{\sigma}}^p  \label{eq:ttildesym}
\end{eqnarray}
with $\xi = ( \xi_1 - \xi_2) / 2 = \pm 1$. We checked numerically that the
inversion parities $\xi_l$ of the two lowest modes are always opposite for the cases
we considered (similar to the two lowest modes in a potential well). Note
that the sign of $\xi$ does not matter for the calculations of the
interference contrast under the assumptions employed in this paper and we
therefore set $\xi = 1$. Using relation (\ref{eq:ttildesym}) and expressing
$\alpha_r^p$ in terms of Majorana operators, we arrive at Eq.~(2)
given in the main paper:
\begin{eqnarray}
  H_{T, I} & = & \sum_{r k \sigma m} c_{r k \sigma} e^{i \varphi / 2} t_{r
  \sigma m} ( \delta_{r L} \gamma_m + \delta_{r R} \gamma_{5 - m}) \nonumber \\ 
  & & + \text{H.c. .}  \label{eq:hti}
\end{eqnarray}
with
\begin{eqnarray}
  & \begin{array}{lllll}
    t_{L \sigma 1} & = & \tfrac{1}{2} ( \tilde{t}_{L \sigma}^+ + \tilde{t}_{L
    \sigma}^-) & = & t_{\sigma 1},\\
    t_{L \sigma 2} & = & \tfrac{- i}{2} ( \tilde{t}_{L \sigma}^+ -
    \tilde{t}_{L \sigma}^-) & = & t_{\sigma 2},
  \end{array} & 
\end{eqnarray}
and $t_{R \sigma m} = ( - 1)^m \xi \sigma t_{L \bar{\sigma} 5 - m}$ (a similar
relation has also been established in Ref. {\cite{Prada17}} where the spin
quantization is rotated).

\subsection{Parametrization of Majorana tunnel couplings}

As mentioned in the main paper, the Majorana couplings can be parametrized in a
simple way by applying a unitary transformation of the spin degree of freedom
in the leads. Introducing $c_{L k \sigma} = U_{\sigma \sigma'} c'_{L k
\sigma'}$ and $t_{\sigma m} = U_{\sigma \sigma'}^{\dag} t'_{\sigma' m}$ in Eq.~(\ref{eq:hti}), we obtain a tunnel Hamiltonian of the same form with $c
\rightarrow c'$ and $t \rightarrow t'$. Defining $U$ in the general form
\begin{eqnarray}
  U & = & \left(\begin{array}{cc}
    e^{i \kappa_{\uparrow}} \cos ( \tau) & e^{i ( - \zeta +
    \kappa_{\uparrow})} \sin ( \tau)\\
    - e^{i ( \zeta + \kappa_{\downarrow})} \sin ( \tau) & e^{i
    \kappa_{\downarrow}} \cos ( \tau)
  \end{array}\right) ,
\end{eqnarray}
we can satisfy the conditions $t_{\uparrow 1}' \in \mathbbm{R}$,
$t_{\downarrow 1}' = 0$, $\arg ( t_{\uparrow 2}') = \arg ( t_{\downarrow 2}')$
by choosing
\begin{eqnarray}
  \tau & = & \text{arctan} ( | t_{\downarrow 1} / t_{\uparrow 1} |), \\
  \zeta & = & \arg ( t_{\downarrow 1} / t_{\uparrow 1}), \\
  \kappa_{\uparrow} & = & - \arg ( t_{\uparrow 1}), \\
  \kappa_{\downarrow} & = & \kappa_{\uparrow} + \arg \left[ \frac{t_{\uparrow
  2} + e^{- i \zeta} \tan ( \tau) t_{\downarrow 2}}{t_{\downarrow 2} - e^{+ i
  \zeta} \tan ( \tau) t_{\uparrow 2}} \right] . 
\end{eqnarray}
With this form of the Majorana couplings $t'_{\sigma m}$, they can be
parametrized as
\begin{eqnarray}
  t_{\uparrow 1}' & = & t \cos ( \lambda), \\
  t_{\downarrow 1}' & = & 0, \\
  t_{\uparrow 2}' & = & t \sin ( \lambda) \cos ( \beta) e^{i \delta}, \\
  t_{\downarrow 2}' & = & t \sin ( \lambda) \sin ( \beta) e^{i \delta} . 
\end{eqnarray}
Omitting the prime from all quantities, we obtain the expressions stated in
the main paper.

\subsection{Extraction procedure for toy model parameters from solutions
for the 2D model}

We next explain how we extract the parameters of the toy model from the
eigenenergies and eigenstates of the 2D model, which are found numerically.

The first step is to consider only the two modes closest to zero energy and
neglect all other modes:
\begin{eqnarray}
  H_{\text{2D}} & \approx & \tfrac{1}{2} \sum_{| l | \leqslant 2}
  \varepsilon_n \beta_l^{\dag} \beta_l \text{ \ = \ } \tfrac{1}{2}
  \pmb{\beta}^{\dag} \cdot \mathcal{H}'_I \cdot \pmb{\beta} ,
  \label{eq:h2dappr}
\end{eqnarray}
with $\mathcal{H}_I' = \tmop{diag} ( \varepsilon_1, \varepsilon_2, -
\varepsilon_1, - \varepsilon_2)$. Matching the eigenenergies
(\ref{eq:energies}) for the toy model to those obtained from the 2D model
fixes the parameters $\Omega$ and $\varepsilon$ to
\begin{eqnarray}
  \Omega & = & ( \varepsilon_2 - \varepsilon_1) / 2, \\
  \varepsilon & = & \sqrt{\varepsilon_1 \varepsilon_2} . 
\end{eqnarray}
To obtain the tunnel amplitudes for the toy model, we rotate the modes
$\beta_l$ such that the matrix representation of $H_{\text{2D}}$ is given by
Eq.~(\ref{eq:hi}). This yields new modes $\tilde{\alpha}_n$ whose wave
function components yield the tunnel couplings as described in Sec.
\ref{sec:ht2d}.

Clearly, one unitary transformation that transforms Eq.~(\ref{eq:h2dappr}) on
the desired form is given by $\widetilde{\pmb{\alpha}} =
\mathcal{U}^{\dag} \pmb{\beta}$; however, this is not the only possible
transformation. The most general transformation includes an additional phase
factor in the definition of the modes $\beta_l$, i.e.,
$\widetilde{\pmb{\alpha}} = \mathcal{U}^{\dag} \mathcal{W}
\pmb{\beta}$, where $\mathcal{W} = \tmop{diag} ( e^{i \varphi_2}, e^{i
\varphi_1}, e^{- i \varphi_1}, e^{- i \varphi_2})$. The transformation
$\mathcal{W}$ leaves the form of Hamiltonian (\ref{eq:h2dappr}) invariant.
Note that the phase factors $\varphi_1$ and $\varphi_2$ correspond to phase
choice for the eigenstates of $\mathcal{H}_{\tmop{BdG}}$ and those can change
randomly from one point to the next when applying a numerical diagonalization
procedure.

The inclusion of the phase factors is important since it influences where the
two modes
\begin{eqnarray}
  \tilde{\alpha}_l ( \varphi_1, \varphi_2) & = & \sum_{\pmb{n},\sigma}
   [ \tilde{u}_{l \pmb{n} \sigma} \psi_{\pmb{n} \sigma} + \tilde{v}_{l \pmb{n} \sigma}
  \psi_{\pmb{n} \sigma}^{\dag}] 
\end{eqnarray}
are localized within the stripe. Here, we use the components $\tilde{u}_{l \pmb{n} \sigma}$ and $\tilde{v}_{l \pmb{n} \sigma}$ of the solutions of the BdG equation in the tight-binding approximation, where   $\pmb{n} = (n_x,n_y)$ denotes the lattice point ($|n_x| \leq N_x/2, n_y \leq N_y/2$). Here, $\psi_{\pmb{n} \sigma}$ denotes the electron field operator with spin $\sigma$ at lattice point $\pmb{n}$. We next introduce a normalized 1D cut of the wave function along the symmetry axis of the stripe, $\tilde{u}_{l n_y \sigma} = \tilde{u}_{l n_x = 0 n_y \sigma} / \sqrt{p_l} $ and $\tilde{v}_{l n_y \sigma} = \tilde{v}_{l n_x = 0 n_y \sigma} / \sqrt{p_l}$ with $p_l = \sum_{n_y, \sigma} [ |\tilde{u}_{l n_x = 0 n_y \sigma}|^2 + |\tilde{v}_{l n_x = 0 n_y \sigma}|^2  ]$. For this 1D cut, we consider the probability to
find a quasiparticle in mode $l$ in the right half of the stripe:
\begin{eqnarray}
  P_l ( \varphi_1, \varphi_2) & = & \sum_{n_y>0, \sigma}
  \nobracket [ | \tilde{u}_{l n_y \sigma} |^2 + | \tilde{v}_{l n_y \sigma} |^2]
  \nobracket . 
\end{eqnarray}
It can be shown that $P_2 ( \varphi_1, \varphi_2) = 1 - P_1 ( \varphi_1,
\varphi_2)$. Using a numerical optimization routine, we chose $\varphi_1$ and
$\varphi_2$ such that $P_1 ( \varphi_1, \varphi_2)$ is minimized.  We emphasize that there is a freedom how to chose the phases $\varphi_1$ and
$\varphi_2$ and this choice is therefore neither an approximation nor does it require any assumptions.
For this
specific choice of the phases, we identify $\tilde{\alpha}_1 ( \varphi_1,
\varphi_2) \equiv \alpha_L$ and $\tilde{\alpha}_2 ( \varphi_1, \varphi_2) =
\alpha_R$.

Up to this point, this procedure contains no other approximation than the
restriction to the two lowest modes. In the toy model, we additionally neglect
the tunnel couplings of mode $\alpha_r$ to lead $\bar{r}$. This is why it is
important to use modes that are maximally localized at the two ends of
the island. In this way, the toy model is a simplification as compared to a
completely general model of a superconducting island with two modes. For the
plots shown in the main paper, we find that the rotated modes with field operators $\tilde{\alpha}_1, \tilde{\alpha}_2$ are well
localized within one half of the stripe and therefore this approximation works
very well.

\section{Details of the calculations}\label{sec:method}

\subsection{Tight-binding calculations}\label{sec:full-model}

For the numerical diagonalization of the Bogoliubov-de Gennes Hamiltonian (5),
we use a tight-binding approach. The details are discussed in
{\cite{Hell17b}}.

\subsection{Transport calculations}

In this Section, we discuss the details of our transport calculations. We
briefly describe our master-equation approach and the assumptions behind it.
We further give important steps for the computation of the rates needed to set
up the master equation and to compute the current.

In terms of the many-body basis $| n, \pmb{\eta} \rangle$ introduced in
Sec. \ref{sec:2dmodel}, the master equation takes the general form
\begin{eqnarray}
  \dot{P}_{\pmb{\eta}} & = & - \left( \sum_{\pmb{\eta}'}
  \Gamma_{\pmb{\eta}'  \pmb{\eta}} \right) P_{\pmb{\eta}} +
  \sum_{\pmb{\eta}'} \Gamma_{\pmb{\eta}  \pmb{\eta}'}
  P_{\pmb{\eta}'} . 
\end{eqnarray}
where $P_{\pmb{\eta}} = \langle n, \pmb{\eta} | \rho_I | n,
\pmb{\eta} \rangle$ is the occupation probability of state $| n,
\pmb{\eta} \rangle$ for the reduced density matrix $\rho_I$ of the
island. We do not take into account off-diagonal elements of the density
matrix (coherences), which is a good approximation when $\Gamma \ll | \varepsilon_{n,
\pmb{\eta}} - \varepsilon_{n, \pmb{\eta}'} |$ for all states
$\pmb{\eta}$, $\pmb{\eta}'$ within each charge state. As mentioned
in the main paper, we focus on the cotunneling regime, i.e., when only one charge
state $n = n_0$ is occupied. We take $n_0$ to be even.  Furthermore, we consider gate voltages where only states with charge $n = n_0 + 1$ need to be
included as virtual intermediate states, while the contribution from states with charge $n = n_0 - 1$ may be neglected.

We solve for the stationary solution $\dot{P}_{\pmb{\eta}}^{\tmop{st}} =
0$ under the constraint $\sum_{\pmb{\eta}}
P_{\pmb{\eta}}^{\tmop{st}} = 1$ and compute the stationary current as
\begin{eqnarray}
  I^{\tmop{st}} & = & \sum_{\pmb{\eta}' \pmb{\eta}} ( \Gamma^{R
  L}_{\pmb{\eta}' \pmb{\eta}} - \Gamma^{L R}_{\pmb{\eta}' 
  \pmb{\eta}}) P^{\tmop{st}}_{\pmb{\eta}} . 
\end{eqnarray}
Here, $\Gamma_{\pmb{\eta}' \pmb{\eta}}^{r' r}$ denotes a cotunneling
process that transfers an electron from lead $r$ to lead $r'$, where
$\Gamma_{\pmb{\eta}' \pmb{\eta}} = \sum_{r r'}
\Gamma_{\pmb{\eta}' \pmb{\eta}}^{r r'}$.

In the case of the toy model, we restrict our calculations to two
modes, which is a good approximation if cotunneling through the quasiparticle continuum can
be neglected. This requires the island to be long (quantization energy $\pi^2
/ 2 m^{\ast} L^2 > \Delta$ provided the island is nearly depleted, $\mu +
E_{\tmop{SO}} \ll \Delta$) or the gap has to be large ($\text{min} (
\varepsilon, \Omega), T, V_b \ll \Delta$).

The tunnel rates are obtained from the T-matrix approach {\cite{Bruus04}},
\begin{eqnarray}
  \frac{\Gamma_{\pmb{\eta}_f \pmb{\eta}_i}}{2 \pi} & = & \sum_{f, i}
  \rho_i \delta ( E_f - E_i) | \langle \pmb{\eta}_f, f | T ( E_i) |
  \pmb{\eta}_i, i \rangle |^2,  \nonumber \\
  & & \label{eq:rates}
\end{eqnarray}
with the T-matrix
\begin{eqnarray}
  T ( E) & = & H_T + H_T \frac{1}{E - H_I - H_0 + i 0_+} H_T \nonumber \\
  & & + \ldots \ .  
  \label{eq:tmatrix}
\end{eqnarray}
In the above expression, $\pmb{\eta}_i$ ($\pmb{\eta}_f$) denotes the
occupations of the island modes in the initial (final) state in charge state $n=0$. Furthermore $i$
($f$) refers to the initial (final) states of the lead, which we we sum over,
weighted by the probability $\rho_i = e^{- \beta H_0} / \tmop{tr} ( e^{- \beta
H_0})$ for initial state $i$ in the grand canonical ensemble. The many-body
energies are given by $E_{\alpha} = \omega_{\alpha} + \varepsilon_{n_0 
\pmb{\eta}_{\alpha}}$, where $H_I | n_0, \pmb{\eta}_{\alpha} \rangle =
\varepsilon_{n_0 \pmb{\eta}_{\alpha}} | n_0, \pmb{\eta}_{\alpha} \rangle$ (see Sec.
\ref{sec:2dmodel}) and $H_0 | \alpha \rangle = \omega_{\alpha} | \alpha
\rangle$ for $\alpha = i, f$. The rates $\Gamma_{\pmb{\eta}'
\pmb{\eta}}^{r' r}$ \new{($r'\neq r$)} are obtained by accounting only for terms $T ( E)
\sim H_T^{r'} \frac{1}{\Delta E} H_T^r$, where $H_T^r$ is the part of the
tunnel Hamiltonian involving lead $r$.

\new{We include in Eq.~(\ref{eq:tmatrix}) terms of $O ( t_D, t^2)$ and neglect higher-order terms, as well as terms of  $O ( t, t_D^2)$ [even though they formally appear in the perturbation expansion of the T-matrix up to $O(H_T^2)$]. The terms $\sim t$ can be omitted because they correspond to sequential electron tunneling processes that are exponentially suppressed in the cotunneling regime.  The effect of terms $\sim t_D^2$ would be to add an additional contribution $\delta I$ to the current. This corresponds to electron-pair tunneling through the reference arm, which does not exhibit a flux dependence as long as higher-order tunneling terms are neglected. An additional contribution $\delta I$ would reduce the interference contrast in the MBS case somewhat but would not change the findings qualitatively. The contribution $\delta I$ can, at least in principle, be made much smaller than the contributions we account for: Note that the largest interference contrast is given when $t_D \sim t^2 / U$, i.e., when the conductances through the two arms are matched. This means that terms $\sim t_D^2$ can be made smaller by scaling down $t_D$ and $t$ while keeping $t_D \sim t^2/U$.}

Evaluating Eq.~(\ref{eq:rates}), we get for the inelastic cotunneling rates
\begin{eqnarray}
  \frac{\Gamma^{r' r}_{\pmb{\eta}_f \pmb{\eta}_i}}{2 \pi} & = &
  \sum_{r r' \sigma \sigma' \pmb{\eta} \pmb{\eta}'} \nu_r \nu_{r'}
  T^{\pmb{\eta} \pmb{\eta}_f \ast}_{r' \sigma'} T_{r
  \sigma}^{\pmb{\eta} \pmb{\eta}_i} T_{r' \sigma'}^{\pmb{\eta}'
  \pmb{\eta}_f} T^{\pmb{\eta}' \pmb{\eta}_i \ast}_{r \sigma} 
  \nonumber\\
  &  & M ( \varepsilon^{1 0}_{\pmb{\eta} \pmb{\eta}_i},
  \varepsilon^{1 0}_{\pmb{\eta}' \pmb{\eta}_i}, \mu_r, \mu_{r'} +
  \varepsilon_{0 \pmb{\eta}_f} - \varepsilon_{0 \pmb{\eta}_i}), 
\end{eqnarray}
and for the elastic cotunneling rates ($\pmb{\eta}_f = \pmb{\eta}_i$)
\begin{eqnarray}
  \frac{\Gamma_{\pmb{\eta}_i \pmb{\eta}_i}}{2 \pi} & = & \sum_{r r'}
  \nu_r \nu_{r'} \{ 2 ( 1 + \tau_{\tmop{sf}}) | t_D |^2 K ( \mu_r, \mu_{r'})
  \nobracket \nonumber\\
  &  & - 2 \tmop{Re} \sum_{\pmb{\eta} \sigma \sigma'} t_D (
  \delta_{\sigma \sigma'} + \tau_{\tmop{sf}} \delta_{\sigma \bar{\sigma}'})
  T^{\pmb{\eta} \pmb{\eta}_i \ast}_{r' \sigma'} T^{\pmb{\eta}
  \pmb{\eta}_i}_{r \sigma} \nonumber\\
  &  & \text{ \ \ \ \ \ \ \ \ \ \ \ \ \ \ } e^{i \pi \Phi / \Phi_0} L (
  \varepsilon^{1 0}_{\pmb{\eta} \pmb{\eta}_i}, \mu_r, \mu_{r'})
  \nonumber\\
  &  & + \sum_{\pmb{\eta} \pmb{\eta}' \sigma \sigma'}
  T^{\pmb{\eta}' \pmb{\eta}_i \ast}_{r' \sigma'} T^{\pmb{\eta}'
  \pmb{\eta}_i}_{r \sigma} T^{\pmb{\eta} \pmb{\eta}_i}_{r'
  \sigma'} T^{\pmb{\eta} \pmb{\eta}_i \ast}_{r \sigma}  \nonumber\\
  &  & \nobracket \text{ \ \ \ \ \ \ \ \ \ \ \ } M ( \varepsilon^{1
  0}_{\pmb{\eta}' \pmb{\eta}_i}, \varepsilon^{1 0}_{\pmb{\eta}
  \pmb{\eta}_i}, \mu_r, \mu_{r'}) \}, 
\end{eqnarray}
with
\begin{eqnarray}
  & & K(\mu_r, \mu_{r'}) \nonumber \\
  & = & \int d \omega f_r ( \omega) ( 1 - f_{r'} (\omega)) \\
  & = & ( \mu_{r'} - \mu_r) b ( \mu_{r'} - \mu_r), \\
  & & L(E, \mu_r, \mu_{r'}) \nonumber \\
  & = & \int d \omega \frac{f_r ( \omega) ( 1 - f_{r'} (\omega))}{\omega - E + i 0_+},  \label{eq:lint}\\
  & & M(E_1, E_2, \mu_r, \mu_{r'}) \nonumber \\
  & = & \int d \omega \frac{f_r ( \omega) ( 1 - f_{r'} ( \omega))}{( \omega - E_1 + i 0_+) ( \omega - E_2 - i 0_+)} . \label{eq:mint}
\end{eqnarray}
with Fermi function $f_r ( \omega) = 1 / ( e^{( \omega - \mu_r) / T} + 1)$ and
the Bose function $b ( \omega) = 1 / ( e^{\omega / T} - 1)$.

To simplify the calculation of the integrals (\ref{eq:lint}) and
(\ref{eq:mint}), we set  $\omega=0$ in the denominators. This is a good approximation when temperature and voltage bias are small compared to energy differences between the island states in different charge sectors  $(T, V_b \ll U)$.
We obtain the approximate expressions
\begin{eqnarray}
  L ( E, \mu_r, \mu_{r'}) & \approx & \frac{K ( \mu_r, \mu_{r'})}{E}, 
  \label{eq:lappr}\\
  M ( E_1, E_2, \mu_r, \mu_{r'}) & \approx & \frac{K ( \mu_r, \mu_{r'})}{E_1
  E_2} .  \label{eq:mappr}
\end{eqnarray}

\subsection{Analytic expression for interference contrast}

We next derive an analytic expression for the maximal interference contrast
for the toy model in the case $\beta = \delta = 0$. We consider here the case
of general bias voltage $V_b$ and obtain the result (4) given in the main paper
in the limit $V_b \rightarrow 0$.

For the toy model, the master equation takes the simple form
\begin{eqnarray}
  \left(\begin{array}{c}
    \dot{P}_+\\
    \dot{P}_-
  \end{array}\right) & = & \left(\begin{array}{cc}
    - \Gamma_{- +} & \Gamma_{+ -}\\
    \Gamma_{- +} & - \Gamma_{+ -}
  \end{array}\right) \left(\begin{array}{c}
    P_+\\
    P_-
  \end{array}\right), 
\end{eqnarray}
where $P_{\eta}$ denotes the occupation probability of state $| n = n_0,
\eta_1 = \eta_2 = ( 1 + \eta) / 2 \rangle$. Note that there are only two
states denoted by $\eta = \pm$ for each charge state because of the
fermion-parity constraint $( - 1)^{\eta_1 + \eta_2} = ( - 1)^{n_0}$. The
stationary solution is simply given by
\begin{eqnarray}
  P_{\pm}^{\tmop{st}} & = & \frac{\Gamma_{\pm \mp}}{\Gamma_{+ -} + \Gamma_{-
  +}} .  \label{eq:pst}
\end{eqnarray}
Assuming deep Coulomb blockade, i.e., $\varepsilon, \Omega \ll U$, we
approximate $E_1 \approx E_2 \approx U$ in Eqs. (\ref{eq:lappr}) and
(\ref{eq:mappr}) and obtain
\begin{eqnarray}
  \Gamma_{\eta_f \eta_i}^{r' r} & = & K^{r' r}_{\eta_f \eta_i}
  \frac{\gamma^2}{2} D^{r r'}_{\eta_f \eta_i} \\
  \Gamma_{\eta_i \eta_i}^{r' r} & = & K^{r' r}_{\eta_i \eta_i} \left\{ ( 1 + |
  \tau_{\tmop{sf}} |^2) \gamma_D \nobracket + \frac{\gamma^2}{2} D^{r
  r'}_{\eta_i \eta_i} \right. \nonumber\\
  &  & \left. - \sqrt{\gamma_D} \gamma \tmop{Re} ( e^{- i \pi \Phi / \Phi_0}
  C^{r r'}_{\eta_i \eta_i}) \right\} 
\end{eqnarray}
with $\gamma_D = 2 \pi | t_D |^2 \nu_L \nu_R$, $\gamma_r = 2 \pi | t_r |^2
\nu_r / U$, $\gamma = \sqrt{\gamma_L \gamma_R}$, and
\begin{eqnarray}
  K^{r' r}_{\eta_f \eta_i} & = & \frac{1}{\pi} K ( \mu_r, \mu_{r'} +
  \varepsilon_{\eta_f \eta_i}), \\
  C^{r r'}_{\eta_f \eta_i} & = & \frac{1}{t_r t_{r'}} \sum_{\eta \sigma
  \sigma'} ( \delta_{\sigma' \sigma} + \tau_{\tmop{sf}} \delta_{\bar{\sigma}'
  \sigma}) T^{\eta_f \eta \ast}_{r' \sigma} T^{\eta \eta_i}_{r \sigma'}, \\
  D^{r r'}_{\eta_f \eta_i} & = & \frac{1}{t_r^2 t^2_{r'}} \sum_{\eta \eta'
  \sigma \sigma'} T^{\eta \eta_f \ast}_{r' \sigma'} T_{r \sigma}^{\eta \eta_i}
  T_{r' \sigma'}^{\eta' \eta_f} T^{\eta' \eta_i \ast}_{r \sigma} . 
\end{eqnarray}
So far, no approximations regarding the tunnel couplings have been made.

We now limit our considerations to the case $\beta = \delta = 0$. By computing
the above sums over the tunnel matrix elements, it is straightforward to show
that $D^{r' r}_{\eta_f \eta_i} = | C^{r' r}_{\eta_f \eta_i} |^2$, $C^{r
r}_{\eta \eta} = 1$, $C^{r r}_{\bar{\eta} \eta} = 0$, $C^{\bar{r} r}_{\eta
\eta} = \cos ( \rho) e^{i \eta \pi / 2}$ and $| C^{\bar{r} r}_{\bar{\eta}
\eta} | = \sin ( \rho)$ with
\begin{eqnarray}
  \cos ( \rho) & = & \frac{\cos ( 2 \lambda)}{\sqrt{( \varepsilon / \Omega)^2
  + 1}} . 
\end{eqnarray}
Inserting the resulting expressions for the rates (\ref{eq:rates}) into the
expression (\ref{eq:pst}) for the stationary occupations yields
\begin{eqnarray}
  P_{\eta}^{\tmop{st}} & = & \frac{G_{\eta}}{\sum_{\eta'} G_{\eta'}}, 
\end{eqnarray}
and the stationary current reads for $\tau_{\text{sf}}=1$
\begin{eqnarray}
  \frac{I_{\tmop{st}}}{V_b} & = & 2 \gamma_D + \frac{\gamma^2}{2} \left[
  \cos^2 ( \rho) + \sin^2 ( \rho) \sum_{\eta} F_{\eta}
  P^{\tmop{st}}_{\bar{\eta}} \right] \nonumber\\
  &  & - \sqrt{\gamma_D} \gamma \cos ( \rho) \sin \left( \pi
  \tfrac{\Phi}{\Phi_0} \right) ( P^{\tmop{st}}_+ - P^{\tmop{st}}_-), 
\end{eqnarray}
with
\begin{eqnarray}
  F_{\eta} & = & \sum_p ( - 1 + 2 p \eta E / V_b) b ( - p V_b + 2 \eta E), \\
  G_{\eta} & = & \sum_p ( - p + 2 \eta E / V_b) b ( - p V_b + 2 \eta E), 
\end{eqnarray}
where $E = \sqrt{\Omega^2 + \varepsilon^2}$. Finally, the maximal interference
contrast reads
\begin{eqnarray}
  \tmop{MIC} & = & \frac{P^{\tmop{st}}_- - P^{\tmop{st}}_+}{2 \sqrt{1 + \left(
  \frac{( \varepsilon / \Omega)^2 + 1}{\cos^2 ( 2 \lambda)} - 1 \right)
  \sum_{\eta} F_{\eta} P^{\tmop{st}}_{\bar{\eta}}}} . 
  \label{eq:micfinitebias}
\end{eqnarray}
In the limit $V_b \rightarrow 0$, one obtains
\begin{eqnarray}
  P_{\eta}^{\tmop{st}} ( V_b = 0) & = & \frac{2 e^{- \eta E / T}}{\cosh ( E /
  T)}, \\
  F_{\eta} ( V_b = 0) & = & \frac{( \eta 2 E / T - 1) + e^{- \eta 2 E / T}}{2
  \sinh^2 ( E / T)}, 
\end{eqnarray}
and inserting this into Eq.~(\ref{eq:micfinitebias}) yields Eq.~(4) in the
main paper.

\section{Discussion of results}\label{sec:method}

\Figure{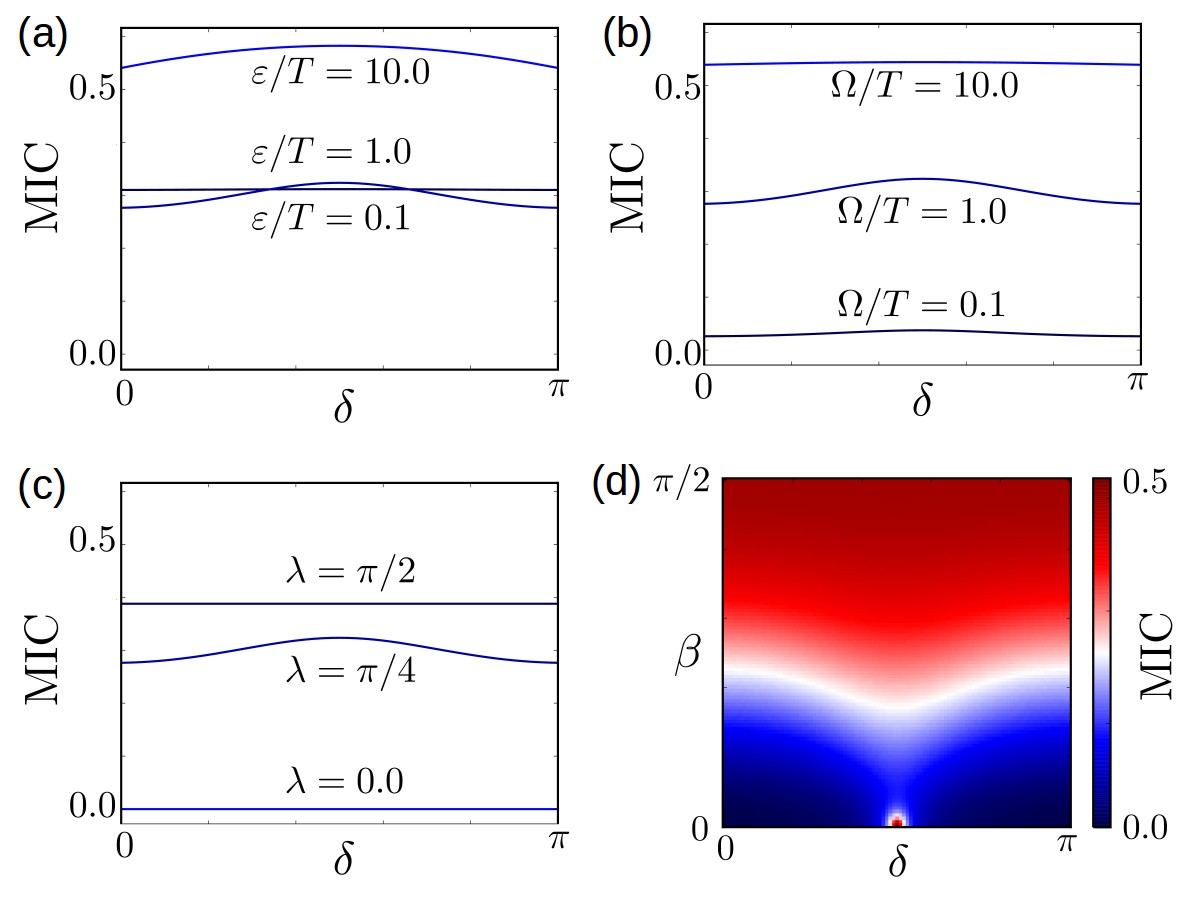}{Dependence
  of the MIC on the spin phase $\delta$ of the Majorana couplings. Except for
  the parameters varied in each panel as indicated, we use the following
  parameters: $\Omega / T = \varepsilon / T = 1$, $\lambda = \beta = \pi / 4$,
  $V_b = 0.01 T$, $\delta = 0$, $U = 100 T$.\label{fig:delta-dep}}

\Figure{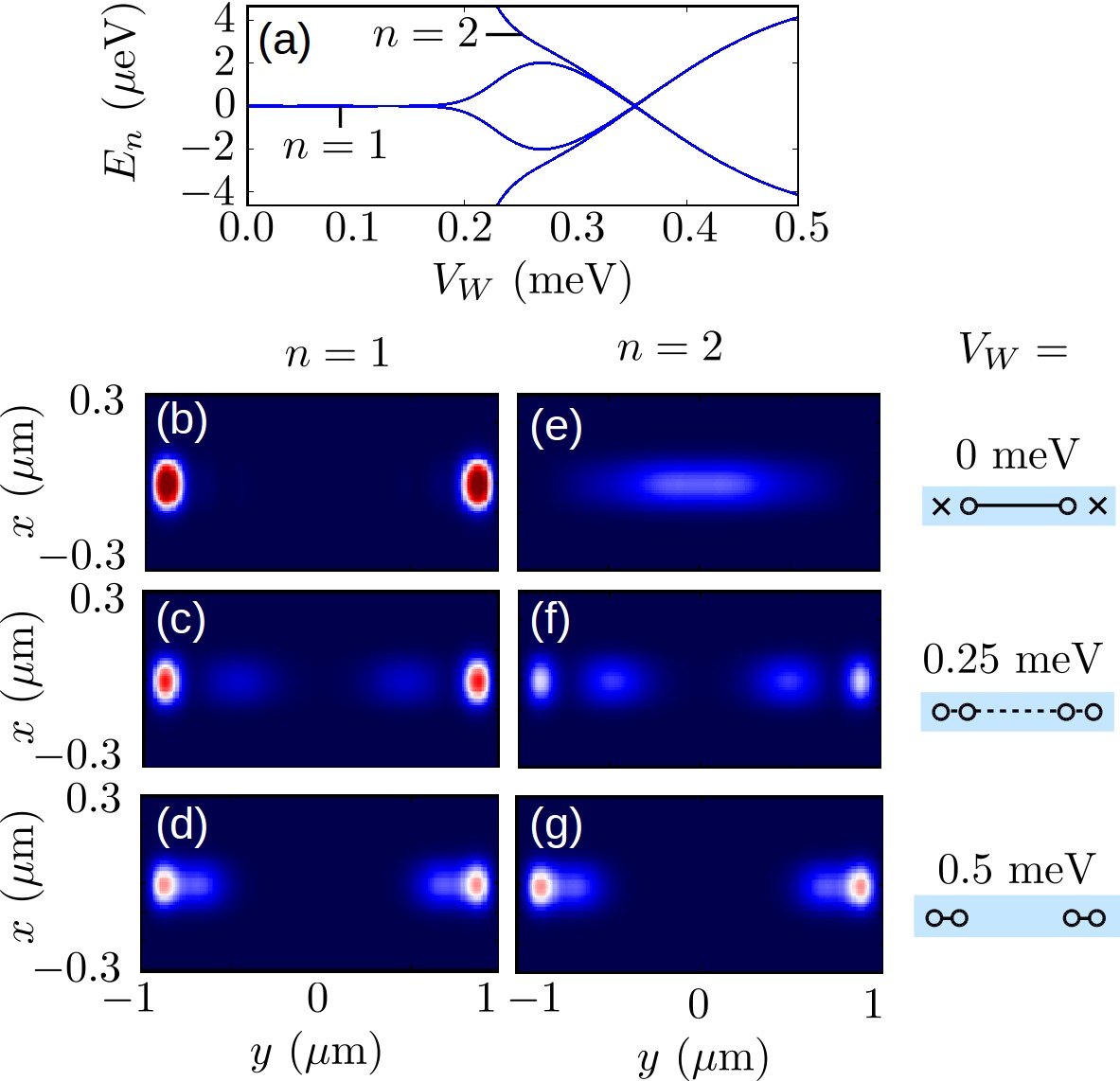}{Tuning
  the probability density of the eigenmodes with the wire potential. In(a), we
  show the same close-up as in Fig. 3(b) of the main paper. In (b) -- (g), we
  show the probability density $P ( n_x, n_y) = \sum_{\sigma}
  [ |u_{\sigma} ( n_x, n_y) |^2+|v_{\sigma} ( n_x, n_y) |^2 ]$  of the lowest eigenmode ($n = 1$, left
  panels) and the next-to-lowest eigenmode ($n = 2$, right panels). The wire
  potential is varied as indicated to the right of the panels with the
  pictograms sketching the corresponding couplings of the Majoranas as in the
  toy model. Parameters are as in Fig. 3, left panel.\label{fig:prob-dens}}

\subsection{Toy model: Pathological case}

As mentioned in the main paper, the case $\Omega = 0$ combined with $\lambda =
0$ or $\varepsilon = 0$ is a pathological case of our model. The reason is
that the mode $\langle 23 \rangle$ decouples from the leads. The occupation
probability of mode $\langle 23 \rangle$ is then not determined by transport
but by other parity-switching mechanisms not included in our model \newer{(e. g. quasiparticle poisoning)}. If this
switching mechanism is much slower than the time window over which the current
is averaged, one may measure a nonzero MIC. 
In this case, the parity of mode
$\langle 23 \rangle$ is fixed during that time and only parity-conserving
tunnel processes happen. \newer{The interference patterns are then shifted by $\pi$ with respect to each other depending on the parity of mode $\langle 23 \rangle$. This can be used to read out the parity of this mode as proposed earlier for Majorana box qubits \cite{Plugge17}.
Furthermore, each
time a transition between ground and excited state happens, the current would switch in experiments (for fixed flux $\Phi$)}.
If the
parity switching is instead fast compared to the current averaging time, the
measurement will average over both patterns and the MIC is zero. The same
reasoning also applies when $\Omega / \varepsilon \rightarrow \infty$ and
parity-flipping cotunneling processes are strongly suppressed.

\subsection{Toy model: Dependence of the maximal interference contrast on
$\delta$}

In Fig. \ref{fig:delta-dep}, we show that the dependence of the interference
contrast on $\delta$ is very weak. The only exception is a spot close to $(
\beta = 0, \delta = \pi / 2)$ \ [Fig. \ref{fig:delta-dep}(d)].

\subsection{2D island model: Tuning from two MBSs to two ABSs}

In this Section, we discuss how the probability densities of the eigenmodes
evolve when increasing the wire potential barrier $V_W$ [Fig.
\ref{fig:prob-dens}]. For $V_W = 0$, there is one mode ($n = 1$) close to zero
energy [Fig. \ref{fig:prob-dens}(a)]. This mode is formed by slightly
overlapping MBSs at opposite ends of the stripe [Fig. \ref{fig:prob-dens}(b)].
The second mode ($n = 2$) is at a large energy $( \approx 0.6 \Delta)$ with a
probability density delocalized along the stripe [Fig.
\ref{fig:prob-dens}(e)]. When increasing $V_W$, this mode comes close to and
sticks to zero energy [Fig. \ref{fig:prob-dens}(a)]. Its probability density
is increasingly pushed to the end of the wire [Fig. \ref{fig:prob-dens}(f)].
For large $V_W$, the two modes become nearly degenerate [Fig.
\ref{fig:prob-dens}(a)] and their probability densities become similar [Fig.
\ref{fig:prob-dens}(d) and (g)]. Here, the eigenstates are symmetric and
antisymmetric combinations of slightly overlapping localized terminal ABSs, so
that the probability densities have equal weights on both ends.

\putbib[cite]

\begin{thebibliography}{79}
\expandafter\ifx\csname natexlab\endcsname\relax\def\natexlab#1{#1}\fi
\expandafter\ifx\csname bibnamefont\endcsname\relax
  \def\bibnamefont#1{#1}\fi
\expandafter\ifx\csname bibfnamefont\endcsname\relax
  \def\bibfnamefont#1{#1}\fi
\expandafter\ifx\csname citenamefont\endcsname\relax
  \def\citenamefont#1{#1}\fi
\expandafter\ifx\csname url\endcsname\relax
  \def\url#1{\texttt{#1}}\fi
\expandafter\ifx\csname urlprefix\endcsname\relax\def\urlprefix{URL }\fi
\providecommand{\bibinfo}[2]{#2}
\providecommand{\eprint}[2][]{\url{#2}}

\bibitem[{\citenamefont{Yu}(1965)}]{Yu65}
\bibinfo{author}{\bibfnamefont{L.}~\bibnamefont{Yu}}, \bibinfo{journal}{Acta
  Physica Sinica} \textbf{\bibinfo{volume}{21}}, \bibinfo{pages}{75}
  (\bibinfo{year}{1965}).

\bibitem[{\citenamefont{Shiba}(1968)}]{Shiba68}
\bibinfo{author}{\bibfnamefont{H.}~\bibnamefont{Shiba}},
  \bibinfo{journal}{Progress of theoretical Physics}
  \textbf{\bibinfo{volume}{40}}, \bibinfo{pages}{435} (\bibinfo{year}{1968}).

\bibitem[{\citenamefont{Rusinov}(1969)}]{Rusinov69}
\bibinfo{author}{\bibfnamefont{A.}~\bibnamefont{Rusinov}},
  \bibinfo{journal}{Soviet Journal of Experimental and Theoretical Physics
  Letters} \textbf{\bibinfo{volume}{9}}, \bibinfo{pages}{85}
  (\bibinfo{year}{1969}).

\bibitem[{\citenamefont{L{\"o}fwander et~al.}(2001)\citenamefont{L{\"o}fwander,
  Shumeiko, and Wendin}}]{Lofwander01}
\bibinfo{author}{\bibfnamefont{T.}~\bibnamefont{L{\"o}fwander}},
  \bibinfo{author}{\bibfnamefont{V.}~\bibnamefont{Shumeiko}}, \bibnamefont{and}
  \bibinfo{author}{\bibfnamefont{G.}~\bibnamefont{Wendin}},
  \bibinfo{journal}{Superconductor Science and Technology}
  \textbf{\bibinfo{volume}{14}}, \bibinfo{pages}{R53} (\bibinfo{year}{2001}).

\bibitem[{\citenamefont{Kulik}(1969)}]{Kulik69}
\bibinfo{author}{\bibfnamefont{I.}~\bibnamefont{Kulik}},
  \bibinfo{journal}{Soviet Journal of Experimental and Theoretical Physics}
  \textbf{\bibinfo{volume}{30}}, \bibinfo{pages}{944} (\bibinfo{year}{1969}).

\bibitem[{\citenamefont{Kitaev}(2001)}]{1DwiresKitaev}
\bibinfo{author}{\bibfnamefont{A.~Y.} \bibnamefont{Kitaev}},
  \bibinfo{journal}{Sov. Phys.--Uspeki} \textbf{\bibinfo{volume}{44}},
  \bibinfo{pages}{131} (\bibinfo{year}{2001}).

\bibitem[{\citenamefont{Alicea}(2012)}]{AliceaReview}
\bibinfo{author}{\bibfnamefont{J.}~\bibnamefont{Alicea}},
  \bibinfo{journal}{Rep. Prog. Phys.} \textbf{\bibinfo{volume}{75}},
  \bibinfo{pages}{076501} (\bibinfo{year}{2012}).

\bibitem[{\citenamefont{Leijnse and Flensberg}(2012)}]{FlensbergReview}
\bibinfo{author}{\bibfnamefont{M.}~\bibnamefont{Leijnse}} \bibnamefont{and}
  \bibinfo{author}{\bibfnamefont{K.}~\bibnamefont{Flensberg}},
  \bibinfo{journal}{Semicond. Sci. Technol.} \textbf{\bibinfo{volume}{27}},
  \bibinfo{pages}{124003} (\bibinfo{year}{2012}).

\bibitem[{\citenamefont{Beenakker}(2013)}]{BeenakkerReview}
\bibinfo{author}{\bibfnamefont{C.~W.~J.} \bibnamefont{Beenakker}},
  \bibinfo{journal}{Annu. Rev. Con. Mat. Phys.} \textbf{\bibinfo{volume}{4}},
  \bibinfo{pages}{113} (\bibinfo{year}{2013}).

\bibitem[{\citenamefont{Stanescu and Tewari}(2013)}]{TewariReview}
\bibinfo{author}{\bibfnamefont{T.~D.} \bibnamefont{Stanescu}} \bibnamefont{and}
  \bibinfo{author}{\bibfnamefont{S.}~\bibnamefont{Tewari}},
  \bibinfo{journal}{J. Phys.: Condens. Matter} \textbf{\bibinfo{volume}{25}},
  \bibinfo{pages}{233201} (\bibinfo{year}{2013}).

\bibitem[{\citenamefont{Alicea et~al.}(2011)\citenamefont{Alicea, Oreg, Refael,
  {von Oppen}, and Fisher}}]{AliceaBraiding}
\bibinfo{author}{\bibfnamefont{J.}~\bibnamefont{Alicea}},
  \bibinfo{author}{\bibfnamefont{Y.}~\bibnamefont{Oreg}},
  \bibinfo{author}{\bibfnamefont{G.}~\bibnamefont{Refael}},
  \bibinfo{author}{\bibfnamefont{F.}~\bibnamefont{{von Oppen}}},
  \bibnamefont{and} \bibinfo{author}{\bibfnamefont{M.~P.~A.}
  \bibnamefont{Fisher}}, \bibinfo{journal}{Nat. Phys.}
  \textbf{\bibinfo{volume}{7}}, \bibinfo{pages}{412} (\bibinfo{year}{2011}).

\bibitem[{\citenamefont{Clarke et~al.}(2011)\citenamefont{Clarke, Sau, and
  Tewari}}]{ClarkeBraiding}
\bibinfo{author}{\bibfnamefont{D.~J.} \bibnamefont{Clarke}},
  \bibinfo{author}{\bibfnamefont{J.~D.} \bibnamefont{Sau}}, \bibnamefont{and}
  \bibinfo{author}{\bibfnamefont{S.}~\bibnamefont{Tewari}},
  \bibinfo{journal}{Phys. Rev. B} \textbf{\bibinfo{volume}{84}},
  \bibinfo{pages}{035120} (\bibinfo{year}{2011}).

\bibitem[{\citenamefont{Halperin et~al.}(2012)\citenamefont{Halperin, Oreg,
  Stern, Refael, Alicea, and von Oppen}}]{HalperinBraiding}
\bibinfo{author}{\bibfnamefont{B.~I.} \bibnamefont{Halperin}},
  \bibinfo{author}{\bibfnamefont{Y.}~\bibnamefont{Oreg}},
  \bibinfo{author}{\bibfnamefont{A.}~\bibnamefont{Stern}},
  \bibinfo{author}{\bibfnamefont{G.}~\bibnamefont{Refael}},
  \bibinfo{author}{\bibfnamefont{J.}~\bibnamefont{Alicea}}, \bibnamefont{and}
  \bibinfo{author}{\bibfnamefont{F.}~\bibnamefont{von Oppen}},
  \bibinfo{journal}{Phys. Rev. B} \textbf{\bibinfo{volume}{85}},
  \bibinfo{pages}{144501} (\bibinfo{year}{2012}).

\bibitem[{\citenamefont{Hyart et~al.}(2013)\citenamefont{Hyart, van Heck,
  Fulga, Burrello, Akhmerov, and Beenakker}}]{BeenakkerBraiding}
\bibinfo{author}{\bibfnamefont{T.}~\bibnamefont{Hyart}},
  \bibinfo{author}{\bibfnamefont{B.}~\bibnamefont{van Heck}},
  \bibinfo{author}{\bibfnamefont{I.~C.} \bibnamefont{Fulga}},
  \bibinfo{author}{\bibfnamefont{M.}~\bibnamefont{Burrello}},
  \bibinfo{author}{\bibfnamefont{A.~R.} \bibnamefont{Akhmerov}},
  \bibnamefont{and} \bibinfo{author}{\bibfnamefont{C.~W.~J.}
  \bibnamefont{Beenakker}}, \bibinfo{journal}{Phys. Rev. B}
  \textbf{\bibinfo{volume}{88}}, \bibinfo{pages}{035121}
  (\bibinfo{year}{2013}).

\bibitem[{\citenamefont{Aasen et~al.}(2016)\citenamefont{Aasen, Hell, Mishmash,
  Higginbotham, Danon, Leijnse, Jespersen, Folk, Marcus, Flensberg
  et~al.}}]{Aasen16}
\bibinfo{author}{\bibfnamefont{D.}~\bibnamefont{Aasen}},
  \bibinfo{author}{\bibfnamefont{M.}~\bibnamefont{Hell}},
  \bibinfo{author}{\bibfnamefont{R.~V.} \bibnamefont{Mishmash}},
  \bibinfo{author}{\bibfnamefont{A.}~\bibnamefont{Higginbotham}},
  \bibinfo{author}{\bibfnamefont{J.}~\bibnamefont{Danon}},
  \bibinfo{author}{\bibfnamefont{M.}~\bibnamefont{Leijnse}},
  \bibinfo{author}{\bibfnamefont{T.~S.} \bibnamefont{Jespersen}},
  \bibinfo{author}{\bibfnamefont{J.~A.} \bibnamefont{Folk}},
  \bibinfo{author}{\bibfnamefont{C.~M.} \bibnamefont{Marcus}},
  \bibinfo{author}{\bibfnamefont{K.}~\bibnamefont{Flensberg}},
  \bibnamefont{et~al.}, \bibinfo{journal}{Phys. Rev. X}
  \textbf{\bibinfo{volume}{6}}, \bibinfo{pages}{031016} (\bibinfo{year}{2016}).

\bibitem[{\citenamefont{Hell et~al.}(2016)\citenamefont{Hell, Danon, Flensberg,
  and Leijnse}}]{Hell16}
\bibinfo{author}{\bibfnamefont{M.}~\bibnamefont{Hell}},
  \bibinfo{author}{\bibfnamefont{J.}~\bibnamefont{Danon}},
  \bibinfo{author}{\bibfnamefont{K.}~\bibnamefont{Flensberg}},
  \bibnamefont{and} \bibinfo{author}{\bibfnamefont{M.}~\bibnamefont{Leijnse}},
  \bibinfo{journal}{Phys. Rev. B} \textbf{\bibinfo{volume}{94}},
  \bibinfo{pages}{035424} (\bibinfo{year}{2016}).

\bibitem[{\citenamefont{Sau et~al.}(2011)\citenamefont{Sau, Clarke, and
  Tewari}}]{SauBraiding}
\bibinfo{author}{\bibfnamefont{J.~D.} \bibnamefont{Sau}},
  \bibinfo{author}{\bibfnamefont{D.~J.} \bibnamefont{Clarke}},
  \bibnamefont{and} \bibinfo{author}{\bibfnamefont{S.}~\bibnamefont{Tewari}},
  \bibinfo{journal}{Phys. Rev. B} \textbf{\bibinfo{volume}{84}},
  \bibinfo{pages}{094505} (\bibinfo{year}{2011}).

\bibitem[{\citenamefont{{van Heck} et~al.}(2012)\citenamefont{{van Heck},
  Akhmerov, Hassler, Burrello, and Beenakker}}]{BraidingWithoutTransport}
\bibinfo{author}{\bibfnamefont{B.}~\bibnamefont{{van Heck}}},
  \bibinfo{author}{\bibfnamefont{A.~R.} \bibnamefont{Akhmerov}},
  \bibinfo{author}{\bibfnamefont{F.}~\bibnamefont{Hassler}},
  \bibinfo{author}{\bibfnamefont{M.}~\bibnamefont{Burrello}}, \bibnamefont{and}
  \bibinfo{author}{\bibfnamefont{C.~W.~J.} \bibnamefont{Beenakker}},
  \bibinfo{journal}{New Journal of Physics} \textbf{\bibinfo{volume}{14}},
  \bibinfo{pages}{035019} (\bibinfo{year}{2012}).

\bibitem[{\citenamefont{Bonderson}(2013)}]{BondersonBraiding}
\bibinfo{author}{\bibfnamefont{P.}~\bibnamefont{Bonderson}},
  \bibinfo{journal}{Phys. Rev. B} \textbf{\bibinfo{volume}{87}},
  \bibinfo{pages}{035113} (\bibinfo{year}{2013}).

\bibitem[{\citenamefont{Vijay and Fu}(2016)}]{Vijay16}
\bibinfo{author}{\bibfnamefont{S.}~\bibnamefont{Vijay}} \bibnamefont{and}
  \bibinfo{author}{\bibfnamefont{L.}~\bibnamefont{Fu}}, \bibinfo{journal}{Phys.
  Rev. B} \textbf{\bibinfo{volume}{94}}, \bibinfo{pages}{235446}
  (\bibinfo{year}{2016}).

\bibitem[{\citenamefont{Karzig et~al.}(2017)\citenamefont{Karzig, Knapp,
  Lutchyn, Bonderson, Hastings, Nayak, Alicea, Flensberg, Plugge, Oreg
  et~al.}}]{Karzig17}
\bibinfo{author}{\bibfnamefont{T.}~\bibnamefont{Karzig}},
  \bibinfo{author}{\bibfnamefont{C.}~\bibnamefont{Knapp}},
  \bibinfo{author}{\bibfnamefont{R.~M.} \bibnamefont{Lutchyn}},
  \bibinfo{author}{\bibfnamefont{P.}~\bibnamefont{Bonderson}},
  \bibinfo{author}{\bibfnamefont{M.~B.} \bibnamefont{Hastings}},
  \bibinfo{author}{\bibfnamefont{C.}~\bibnamefont{Nayak}},
  \bibinfo{author}{\bibfnamefont{J.}~\bibnamefont{Alicea}},
  \bibinfo{author}{\bibfnamefont{K.}~\bibnamefont{Flensberg}},
  \bibinfo{author}{\bibfnamefont{S.}~\bibnamefont{Plugge}},
  \bibinfo{author}{\bibfnamefont{Y.}~\bibnamefont{Oreg}}, \bibnamefont{et~al.},
  \bibinfo{journal}{Phys. Rev. B} \textbf{\bibinfo{volume}{95}},
  \bibinfo{pages}{235305} (\bibinfo{year}{2017}).

\bibitem[{\citenamefont{Bravyi and Kitaev}(2002)}]{BravyiKitaev}
\bibinfo{author}{\bibfnamefont{S.}~\bibnamefont{Bravyi}} \bibnamefont{and}
  \bibinfo{author}{\bibfnamefont{A.}~\bibnamefont{Kitaev}},
  \bibinfo{journal}{Annals of Physics} \textbf{\bibinfo{volume}{298}},
  \bibinfo{pages}{210} (\bibinfo{year}{2002}).

\bibitem[{\citenamefont{Bravyi and Kitaev}(2005)}]{BravyiKitaev2}
\bibinfo{author}{\bibfnamefont{S.}~\bibnamefont{Bravyi}} \bibnamefont{and}
  \bibinfo{author}{\bibfnamefont{A.}~\bibnamefont{Kitaev}},
  \bibinfo{journal}{Phys.\ Rev.\ A} \textbf{\bibinfo{volume}{71}},
  \bibinfo{pages}{022316} (\bibinfo{year}{2005}).

\bibitem[{\citenamefont{Freedman et~al.}(2003)\citenamefont{Freedman, Kitaev,
  Larsen, and Wang}}]{Freedman03}
\bibinfo{author}{\bibfnamefont{M.~H.} \bibnamefont{Freedman}},
  \bibinfo{author}{\bibfnamefont{A.}~\bibnamefont{Kitaev}},
  \bibinfo{author}{\bibfnamefont{M.~J.} \bibnamefont{Larsen}},
  \bibnamefont{and} \bibinfo{author}{\bibfnamefont{Z.}~\bibnamefont{Wang}},
  \bibinfo{journal}{Bull.\ Amer.\ Math.\ Soc.} \textbf{\bibinfo{volume}{40}},
  \bibinfo{pages}{31} (\bibinfo{year}{2003}).

\bibitem[{\citenamefont{Oreg et~al.}(2010)\citenamefont{Oreg, Refael, and {von
  Oppen}}}]{1DwiresOreg}
\bibinfo{author}{\bibfnamefont{Y.}~\bibnamefont{Oreg}},
  \bibinfo{author}{\bibfnamefont{G.}~\bibnamefont{Refael}}, \bibnamefont{and}
  \bibinfo{author}{\bibfnamefont{F.}~\bibnamefont{{von Oppen}}},
  \bibinfo{journal}{Phys.\ Rev.\ Lett.} \textbf{\bibinfo{volume}{105}},
  \bibinfo{pages}{177002} (\bibinfo{year}{2010}).

\bibitem[{\citenamefont{Lutchyn et~al.}(2010)\citenamefont{Lutchyn, Sau, and
  Das~Sarma}}]{1DwiresLutchyn}
\bibinfo{author}{\bibfnamefont{R.~M.} \bibnamefont{Lutchyn}},
  \bibinfo{author}{\bibfnamefont{J.~D.} \bibnamefont{Sau}}, \bibnamefont{and}
  \bibinfo{author}{\bibfnamefont{S.}~\bibnamefont{Das~Sarma}},
  \bibinfo{journal}{Phys.\ Rev.\ Lett.} \textbf{\bibinfo{volume}{105}},
  \bibinfo{pages}{077001} (\bibinfo{year}{2010}).

\bibitem[{\citenamefont{Mourik et~al.}(2012)\citenamefont{Mourik, Zuo, Frolov,
  Plissard, Bakkers, and Kouwenhoven}}]{mourik12}
\bibinfo{author}{\bibfnamefont{V.}~\bibnamefont{Mourik}},
  \bibinfo{author}{\bibfnamefont{K.}~\bibnamefont{Zuo}},
  \bibinfo{author}{\bibfnamefont{S.~M.} \bibnamefont{Frolov}},
  \bibinfo{author}{\bibfnamefont{S.~R.} \bibnamefont{Plissard}},
  \bibinfo{author}{\bibfnamefont{E.~P. A.~M.} \bibnamefont{Bakkers}},
  \bibnamefont{and} \bibinfo{author}{\bibfnamefont{L.~P.}
  \bibnamefont{Kouwenhoven}}, \bibinfo{journal}{Science}
  \textbf{\bibinfo{volume}{336}}, \bibinfo{pages}{1003} (\bibinfo{year}{2012}).

\bibitem[{\citenamefont{Das et~al.}(2012)\citenamefont{Das, Ronen, Most, Oreg,
  Heiblum, and Shtrikman}}]{das12}
\bibinfo{author}{\bibfnamefont{A.}~\bibnamefont{Das}},
  \bibinfo{author}{\bibfnamefont{Y.}~\bibnamefont{Ronen}},
  \bibinfo{author}{\bibfnamefont{Y.}~\bibnamefont{Most}},
  \bibinfo{author}{\bibfnamefont{Y.}~\bibnamefont{Oreg}},
  \bibinfo{author}{\bibfnamefont{M.}~\bibnamefont{Heiblum}}, \bibnamefont{and}
  \bibinfo{author}{\bibfnamefont{H.}~\bibnamefont{Shtrikman}},
  \bibinfo{journal}{Nat. Phys.} \textbf{\bibinfo{volume}{8}},
  \bibinfo{pages}{887} (\bibinfo{year}{2012}).

\bibitem[{\citenamefont{Finck et~al.}(2013)\citenamefont{Finck, Van~Harlingen,
  Mohseni, Jung, and Li}}]{finck12}
\bibinfo{author}{\bibfnamefont{A.~D.~K.} \bibnamefont{Finck}},
  \bibinfo{author}{\bibfnamefont{D.~J.} \bibnamefont{Van~Harlingen}},
  \bibinfo{author}{\bibfnamefont{P.~K.} \bibnamefont{Mohseni}},
  \bibinfo{author}{\bibfnamefont{K.}~\bibnamefont{Jung}}, \bibnamefont{and}
  \bibinfo{author}{\bibfnamefont{X.}~\bibnamefont{Li}}, \bibinfo{journal}{Phys.
  Rev. Lett.} \textbf{\bibinfo{volume}{110}}, \bibinfo{pages}{126406}
  (\bibinfo{year}{2013}).

\bibitem[{\citenamefont{Rokhinson et~al.}(2012)\citenamefont{Rokhinson, Liu,
  and Furdyna}}]{Rokhinson}
\bibinfo{author}{\bibfnamefont{L.~P.} \bibnamefont{Rokhinson}},
  \bibinfo{author}{\bibfnamefont{X.}~\bibnamefont{Liu}}, \bibnamefont{and}
  \bibinfo{author}{\bibfnamefont{J.~K.} \bibnamefont{Furdyna}},
  \bibinfo{journal}{Nat. Phys.} \textbf{\bibinfo{volume}{8}},
  \bibinfo{pages}{795} (\bibinfo{year}{2012}).

\bibitem[{\citenamefont{Deng et~al.}(2012)\citenamefont{Deng, Yu, Huang,
  Larsson, Caroff, and Xu}}]{Deng12}
\bibinfo{author}{\bibfnamefont{M.~T.} \bibnamefont{Deng}},
  \bibinfo{author}{\bibfnamefont{C.~L.} \bibnamefont{Yu}},
  \bibinfo{author}{\bibfnamefont{G.~Y.} \bibnamefont{Huang}},
  \bibinfo{author}{\bibfnamefont{M.}~\bibnamefont{Larsson}},
  \bibinfo{author}{\bibfnamefont{P.}~\bibnamefont{Caroff}}, \bibnamefont{and}
  \bibinfo{author}{\bibfnamefont{H.~Q.} \bibnamefont{Xu}},
  \bibinfo{journal}{Nano Lett.} \textbf{\bibinfo{volume}{12}},
  \bibinfo{pages}{6414} (\bibinfo{year}{2012}).

\bibitem[{\citenamefont{Churchill et~al.}(2013)\citenamefont{Churchill, Fatemi,
  Grove-Rasmussen, Deng, Caroff, Xu, and Marcus}}]{Churchill13}
\bibinfo{author}{\bibfnamefont{H.~O.~H.} \bibnamefont{Churchill}},
  \bibinfo{author}{\bibfnamefont{V.}~\bibnamefont{Fatemi}},
  \bibinfo{author}{\bibfnamefont{K.}~\bibnamefont{Grove-Rasmussen}},
  \bibinfo{author}{\bibfnamefont{M.~T.} \bibnamefont{Deng}},
  \bibinfo{author}{\bibfnamefont{P.}~\bibnamefont{Caroff}},
  \bibinfo{author}{\bibfnamefont{H.~Q.} \bibnamefont{Xu}}, \bibnamefont{and}
  \bibinfo{author}{\bibfnamefont{C.~M.} \bibnamefont{Marcus}},
  \bibinfo{journal}{Phys. Rev. B} \textbf{\bibinfo{volume}{87}},
  \bibinfo{pages}{241401} (\bibinfo{year}{2013}).

\bibitem[{\citenamefont{Albrecht et~al.}(2016)\citenamefont{Albrecht,
  Higginbotham, Madsen, Kuemmeth, Jespersen, Nyg, Krogstrup, and
  Marcus}}]{Albrecht16}
\bibinfo{author}{\bibfnamefont{S.~M.} \bibnamefont{Albrecht}},
  \bibinfo{author}{\bibfnamefont{A.~P.} \bibnamefont{Higginbotham}},
  \bibinfo{author}{\bibfnamefont{M.}~\bibnamefont{Madsen}},
  \bibinfo{author}{\bibfnamefont{F.}~\bibnamefont{Kuemmeth}},
  \bibinfo{author}{\bibfnamefont{T.~S.} \bibnamefont{Jespersen}},
  \bibinfo{author}{\bibfnamefont{J.}~\bibnamefont{Nyg}},
  \bibinfo{author}{\bibfnamefont{P.}~\bibnamefont{Krogstrup}},
  \bibnamefont{and} \bibinfo{author}{\bibfnamefont{C.~M.}
  \bibnamefont{Marcus}}, \bibinfo{journal}{Nature}
  \textbf{\bibinfo{volume}{531}}, \bibinfo{pages}{206} (\bibinfo{year}{2016}).

\bibitem[{\citenamefont{Deng et~al.}(2016)\citenamefont{Deng, Vaitiek{\.e}nas,
  Hansen, Danon, Leijnse, Flensberg, Nyg{\aa}rd, Krogstrup, and
  Marcus}}]{Deng16}
\bibinfo{author}{\bibfnamefont{M.}~\bibnamefont{Deng}},
  \bibinfo{author}{\bibfnamefont{S.}~\bibnamefont{Vaitiek{\.e}nas}},
  \bibinfo{author}{\bibfnamefont{E.~B.} \bibnamefont{Hansen}},
  \bibinfo{author}{\bibfnamefont{J.}~\bibnamefont{Danon}},
  \bibinfo{author}{\bibfnamefont{M.}~\bibnamefont{Leijnse}},
  \bibinfo{author}{\bibfnamefont{K.}~\bibnamefont{Flensberg}},
  \bibinfo{author}{\bibfnamefont{J.}~\bibnamefont{Nyg{\aa}rd}},
  \bibinfo{author}{\bibfnamefont{P.}~\bibnamefont{Krogstrup}},
  \bibnamefont{and} \bibinfo{author}{\bibfnamefont{C.~M.}
  \bibnamefont{Marcus}}, \bibinfo{journal}{Science}
  \textbf{\bibinfo{volume}{354}}, \bibinfo{pages}{1557} (\bibinfo{year}{2016}).

\bibitem[{\citenamefont{Suominen et~al.}(2017)\citenamefont{Suominen,
  Kjaergaard, Hamilton, Shabani, Palmstr{\o}m, Marcus, and
  Nichele}}]{Suominen17}
\bibinfo{author}{\bibfnamefont{H.~J.} \bibnamefont{Suominen}},
  \bibinfo{author}{\bibfnamefont{M.}~\bibnamefont{Kjaergaard}},
  \bibinfo{author}{\bibfnamefont{A.~R.} \bibnamefont{Hamilton}},
  \bibinfo{author}{\bibfnamefont{J.}~\bibnamefont{Shabani}},
  \bibinfo{author}{\bibfnamefont{C.~J.} \bibnamefont{Palmstr{\o}m}},
  \bibinfo{author}{\bibfnamefont{C.~M.} \bibnamefont{Marcus}},
  \bibnamefont{and} \bibinfo{author}{\bibfnamefont{F.}~\bibnamefont{Nichele}},
  \bibinfo{journal}{arXiv preprint arXiv:1703.03699v1}  (\bibinfo{year}{2017}).

\bibitem[{\citenamefont{Nichele et~al.}(2017)\citenamefont{Nichele, Drachmann,
  Whiticar, O'Farrell, Suominen, Fornieri, Wang, Gardner, Thomas, Hatke
  et~al.}}]{Nichele17}
\bibinfo{author}{\bibfnamefont{F.}~\bibnamefont{Nichele}},
  \bibinfo{author}{\bibfnamefont{A.~C.} \bibnamefont{Drachmann}},
  \bibinfo{author}{\bibfnamefont{A.~M.} \bibnamefont{Whiticar}},
  \bibinfo{author}{\bibfnamefont{E.~C.} \bibnamefont{O'Farrell}},
  \bibinfo{author}{\bibfnamefont{H.~J.} \bibnamefont{Suominen}},
  \bibinfo{author}{\bibfnamefont{A.}~\bibnamefont{Fornieri}},
  \bibinfo{author}{\bibfnamefont{T.}~\bibnamefont{Wang}},
  \bibinfo{author}{\bibfnamefont{G.~C.} \bibnamefont{Gardner}},
  \bibinfo{author}{\bibfnamefont{C.}~\bibnamefont{Thomas}},
  \bibinfo{author}{\bibfnamefont{A.~T.} \bibnamefont{Hatke}},
  \bibnamefont{et~al.}, \bibinfo{journal}{Phys. Rev. Lett.}
  \textbf{\bibinfo{volume}{119}}, \bibinfo{pages}{136803}
  (\bibinfo{year}{2017}).

\bibitem[{\citenamefont{Bolech and Demler}(2007)}]{ZeroBiasAnomaly1}
\bibinfo{author}{\bibfnamefont{C.~J.} \bibnamefont{Bolech}} \bibnamefont{and}
  \bibinfo{author}{\bibfnamefont{E.}~\bibnamefont{Demler}},
  \bibinfo{journal}{Phys. Rev. Lett.} \textbf{\bibinfo{volume}{98}},
  \bibinfo{pages}{237002} (\bibinfo{year}{2007}).

\bibitem[{\citenamefont{Law et~al.}(2009)\citenamefont{Law, Lee, and
  Ng}}]{ZeroBiasAnomaly3}
\bibinfo{author}{\bibfnamefont{K.~T.} \bibnamefont{Law}},
  \bibinfo{author}{\bibfnamefont{P.~A.} \bibnamefont{Lee}}, \bibnamefont{and}
  \bibinfo{author}{\bibfnamefont{T.~K.} \bibnamefont{Ng}},
  \bibinfo{journal}{Phys.\ Rev.\ Lett.} \textbf{\bibinfo{volume}{103}},
  \bibinfo{pages}{237001} (\bibinfo{year}{2009}).

\bibitem[{\citenamefont{Flensberg}(2010)}]{ZeroBiasAnomaly4}
\bibinfo{author}{\bibfnamefont{K.}~\bibnamefont{Flensberg}},
  \bibinfo{journal}{Phys. Rev. B} \textbf{\bibinfo{volume}{82}},
  \bibinfo{pages}{180516} (\bibinfo{year}{2010}).

\bibitem[{\citenamefont{Golub and Horovitz}(2011)}]{ZeroBiasAnomaly5}
\bibinfo{author}{\bibfnamefont{A.}~\bibnamefont{Golub}} \bibnamefont{and}
  \bibinfo{author}{\bibfnamefont{B.}~\bibnamefont{Horovitz}},
  \bibinfo{journal}{Phys. Rev. B} \textbf{\bibinfo{volume}{83}},
  \bibinfo{pages}{153415} (\bibinfo{year}{2011}).

\bibitem[{\citenamefont{Wimmer et~al.}(2011)\citenamefont{Wimmer, Akhmerov,
  Dahlhaus, and Beenakker}}]{ZeroBiasAnomaly6}
\bibinfo{author}{\bibfnamefont{M.}~\bibnamefont{Wimmer}},
  \bibinfo{author}{\bibfnamefont{A.~R.} \bibnamefont{Akhmerov}},
  \bibinfo{author}{\bibfnamefont{J.~P.} \bibnamefont{Dahlhaus}},
  \bibnamefont{and} \bibinfo{author}{\bibfnamefont{C.~W.~J.}
  \bibnamefont{Beenakker}}, \bibinfo{journal}{New J. Phys.}
  \textbf{\bibinfo{volume}{13}}, \bibinfo{pages}{053016}
  (\bibinfo{year}{2011}).

\bibitem[{\citenamefont{Bagrets and Altland}(2012)}]{MajoranaImposter2}
\bibinfo{author}{\bibfnamefont{D.}~\bibnamefont{Bagrets}} \bibnamefont{and}
  \bibinfo{author}{\bibfnamefont{A.}~\bibnamefont{Altland}},
  \bibinfo{journal}{Phys. Rev. Lett.} \textbf{\bibinfo{volume}{109}},
  \bibinfo{pages}{227005} (\bibinfo{year}{2012}).

\bibitem[{\citenamefont{Liu et~al.}(2012)\citenamefont{Liu, Potter, Law, and
  Lee}}]{MajoranaImposter3}
\bibinfo{author}{\bibfnamefont{J.}~\bibnamefont{Liu}},
  \bibinfo{author}{\bibfnamefont{A.~C.} \bibnamefont{Potter}},
  \bibinfo{author}{\bibfnamefont{K.~T.} \bibnamefont{Law}}, \bibnamefont{and}
  \bibinfo{author}{\bibfnamefont{P.~A.} \bibnamefont{Lee}},
  \bibinfo{journal}{Phys. Rev. Lett.} \textbf{\bibinfo{volume}{109}},
  \bibinfo{pages}{267002} (\bibinfo{year}{2012}).

\bibitem[{\citenamefont{Pikulin et~al.}(2012)\citenamefont{Pikulin, Dahlhaus,
  Wimmer, Schomerus, and Beenakker}}]{MajoranaImposter4}
\bibinfo{author}{\bibfnamefont{D.~I.} \bibnamefont{Pikulin}},
  \bibinfo{author}{\bibfnamefont{J.~P.} \bibnamefont{Dahlhaus}},
  \bibinfo{author}{\bibfnamefont{M.}~\bibnamefont{Wimmer}},
  \bibinfo{author}{\bibfnamefont{H.}~\bibnamefont{Schomerus}},
  \bibnamefont{and} \bibinfo{author}{\bibfnamefont{C.~W.~J.}
  \bibnamefont{Beenakker}}, \bibinfo{journal}{New Journal of Physics}
  \textbf{\bibinfo{volume}{14}}, \bibinfo{pages}{125011}
  (\bibinfo{year}{2012}).

\bibitem[{\citenamefont{Goldhaber-Gordon
  et~al.}(1998)\citenamefont{Goldhaber-Gordon, Shtrikman, Mahalu,
  Abusch-Magder, Meirav, and Kastner}}]{Goldhaber98}
\bibinfo{author}{\bibfnamefont{D.}~\bibnamefont{Goldhaber-Gordon}},
  \bibinfo{author}{\bibfnamefont{H.}~\bibnamefont{Shtrikman}},
  \bibinfo{author}{\bibfnamefont{D.}~\bibnamefont{Mahalu}},
  \bibinfo{author}{\bibfnamefont{D.}~\bibnamefont{Abusch-Magder}},
  \bibinfo{author}{\bibfnamefont{U.}~\bibnamefont{Meirav}}, \bibnamefont{and}
  \bibinfo{author}{\bibfnamefont{M.}~\bibnamefont{Kastner}},
  \bibinfo{journal}{Nature} \textbf{\bibinfo{volume}{391}}
  (\bibinfo{year}{1998}).

\bibitem[{\citenamefont{Lee et~al.}(2012)\citenamefont{Lee, Jiang, Aguado,
  Katsaros, Lieber, and De~Franceschi}}]{MajoranaImposter1}
\bibinfo{author}{\bibfnamefont{E.~J.~H.} \bibnamefont{Lee}},
  \bibinfo{author}{\bibfnamefont{X.}~\bibnamefont{Jiang}},
  \bibinfo{author}{\bibfnamefont{R.}~\bibnamefont{Aguado}},
  \bibinfo{author}{\bibfnamefont{G.}~\bibnamefont{Katsaros}},
  \bibinfo{author}{\bibfnamefont{C.~M.} \bibnamefont{Lieber}},
  \bibnamefont{and}
  \bibinfo{author}{\bibfnamefont{S.}~\bibnamefont{De~Franceschi}},
  \bibinfo{journal}{Phys. Rev. Lett.} \textbf{\bibinfo{volume}{109}},
  \bibinfo{pages}{186802} (\bibinfo{year}{2012}).

\bibitem[{\citenamefont{Kells et~al.}(2012{\natexlab{a}})\citenamefont{Kells,
  Meidan, and Brouwer}}]{MajoranaImposter5}
\bibinfo{author}{\bibfnamefont{G.}~\bibnamefont{Kells}},
  \bibinfo{author}{\bibfnamefont{D.}~\bibnamefont{Meidan}}, \bibnamefont{and}
  \bibinfo{author}{\bibfnamefont{P.~W.} \bibnamefont{Brouwer}},
  \bibinfo{journal}{Phys. Rev. B} \textbf{\bibinfo{volume}{86}},
  \bibinfo{pages}{100503} (\bibinfo{year}{2012}{\natexlab{a}}).

\bibitem[{\citenamefont{{Krogstrup} et~al.}(2015)\citenamefont{{Krogstrup},
  {Ziino}, {Chang}, {Albrecht}, {Madsen}, {Johnson}, {Nyg{\aa}rd}, {Marcus},
  and {Jespersen}}}]{Marcus15_NatureMat_14_400}
\bibinfo{author}{\bibfnamefont{P.}~\bibnamefont{{Krogstrup}}},
  \bibinfo{author}{\bibfnamefont{N.~L.~B.} \bibnamefont{{Ziino}}},
  \bibinfo{author}{\bibfnamefont{W.}~\bibnamefont{{Chang}}},
  \bibinfo{author}{\bibfnamefont{S.~M.} \bibnamefont{{Albrecht}}},
  \bibinfo{author}{\bibfnamefont{M.~H.} \bibnamefont{{Madsen}}},
  \bibinfo{author}{\bibfnamefont{E.}~\bibnamefont{{Johnson}}},
  \bibinfo{author}{\bibfnamefont{J.}~\bibnamefont{{Nyg{\aa}rd}}},
  \bibinfo{author}{\bibfnamefont{C.~M.} \bibnamefont{{Marcus}}},
  \bibnamefont{and} \bibinfo{author}{\bibfnamefont{T.~S.}
  \bibnamefont{{Jespersen}}}, \bibinfo{journal}{Nature Materials}
  \textbf{\bibinfo{volume}{14}}, \bibinfo{pages}{400} (\bibinfo{year}{2015}).

\bibitem[{\citenamefont{{Chang} et~al.}(2015)\citenamefont{{Chang}, {Albrecht},
  {Jespersen}, {Kuemmeth}, {Krogstrup}, {Nyg{\aa}rd}, and
  {Marcus}}}]{Marcus15_NatureNano_10_232}
\bibinfo{author}{\bibfnamefont{W.}~\bibnamefont{{Chang}}},
  \bibinfo{author}{\bibfnamefont{S.~M.} \bibnamefont{{Albrecht}}},
  \bibinfo{author}{\bibfnamefont{T.~S.} \bibnamefont{{Jespersen}}},
  \bibinfo{author}{\bibfnamefont{F.}~\bibnamefont{{Kuemmeth}}},
  \bibinfo{author}{\bibfnamefont{P.}~\bibnamefont{{Krogstrup}}},
  \bibinfo{author}{\bibfnamefont{J.}~\bibnamefont{{Nyg{\aa}rd}}},
  \bibnamefont{and} \bibinfo{author}{\bibfnamefont{C.~M.}
  \bibnamefont{{Marcus}}}, \bibinfo{journal}{Nature Nanotechnology}
  \textbf{\bibinfo{volume}{10}}, \bibinfo{pages}{232} (\bibinfo{year}{2015}).

\bibitem[{\citenamefont{Zhang et~al.}(2017)\citenamefont{Zhang, G{\"u}l,
  Conesa-Boj, Nowak, Wimmer, Zuo, Mourik, de~Vries, van Veen, de~Moor
  et~al.}}]{Zhang17}
\bibinfo{author}{\bibfnamefont{H.}~\bibnamefont{Zhang}},
  \bibinfo{author}{\bibfnamefont{{\"O}.}~\bibnamefont{G{\"u}l}},
  \bibinfo{author}{\bibfnamefont{S.}~\bibnamefont{Conesa-Boj}},
  \bibinfo{author}{\bibfnamefont{M.~P.} \bibnamefont{Nowak}},
  \bibinfo{author}{\bibfnamefont{M.}~\bibnamefont{Wimmer}},
  \bibinfo{author}{\bibfnamefont{K.}~\bibnamefont{Zuo}},
  \bibinfo{author}{\bibfnamefont{V.}~\bibnamefont{Mourik}},
  \bibinfo{author}{\bibfnamefont{F.~K.} \bibnamefont{de~Vries}},
  \bibinfo{author}{\bibfnamefont{J.}~\bibnamefont{van Veen}},
  \bibinfo{author}{\bibfnamefont{M.~W.} \bibnamefont{de~Moor}},
  \bibnamefont{et~al.}, \bibinfo{journal}{Nature Communications}
  \textbf{\bibinfo{volume}{8}} (\bibinfo{year}{2017}).

\bibitem[{\citenamefont{Gazibegovic et~al.}(2017)\citenamefont{Gazibegovic,
  Car, Zhang, Balk, Logan, de~Moor, Cassidy, Schmits, Xu, Wang
  et~al.}}]{Gazibegovic17}
\bibinfo{author}{\bibfnamefont{S.}~\bibnamefont{Gazibegovic}},
  \bibinfo{author}{\bibfnamefont{D.}~\bibnamefont{Car}},
  \bibinfo{author}{\bibfnamefont{H.}~\bibnamefont{Zhang}},
  \bibinfo{author}{\bibfnamefont{S.~C.} \bibnamefont{Balk}},
  \bibinfo{author}{\bibfnamefont{J.~A.} \bibnamefont{Logan}},
  \bibinfo{author}{\bibfnamefont{M.~W.} \bibnamefont{de~Moor}},
  \bibinfo{author}{\bibfnamefont{M.~C.} \bibnamefont{Cassidy}},
  \bibinfo{author}{\bibfnamefont{R.}~\bibnamefont{Schmits}},
  \bibinfo{author}{\bibfnamefont{D.}~\bibnamefont{Xu}},
  \bibinfo{author}{\bibfnamefont{G.}~\bibnamefont{Wang}}, \bibnamefont{et~al.},
  \bibinfo{journal}{Nature} \textbf{\bibinfo{volume}{548}}
  (\bibinfo{year}{2017}).

\bibitem[{\citenamefont{Gül et~al.}(2017)\citenamefont{Gül, Zhang,
  de~Vries, van Veen, Zuo, Mourik, Conesa-Boj, Nowak, Van~Woerkom,
  Quintero-P{\'e}rez et~al.}}]{Gul17}
\bibinfo{author}{\bibfnamefont{O.}~\bibnamefont{Gül}},
  \bibinfo{author}{\bibfnamefont{H.}~\bibnamefont{Zhang}},
  \bibinfo{author}{\bibfnamefont{F.~K.} \bibnamefont{de~Vries}},
  \bibinfo{author}{\bibfnamefont{J.}~\bibnamefont{van Veen}},
  \bibinfo{author}{\bibfnamefont{K.}~\bibnamefont{Zuo}},
  \bibinfo{author}{\bibfnamefont{V.}~\bibnamefont{Mourik}},
  \bibinfo{author}{\bibfnamefont{S.}~\bibnamefont{Conesa-Boj}},
  \bibinfo{author}{\bibfnamefont{M.~P.} \bibnamefont{Nowak}},
  \bibinfo{author}{\bibfnamefont{D.~J.} \bibnamefont{Van~Woerkom}},
  \bibinfo{author}{\bibfnamefont{M.}~\bibnamefont{Quintero-P{\'e}rez}},
  \bibnamefont{et~al.}, \bibinfo{journal}{Nano Letters}
  \textbf{\bibinfo{volume}{17}}, \bibinfo{pages}{2690} (\bibinfo{year}{2017}).

\bibitem[{\citenamefont{Kells et~al.}(2012{\natexlab{b}})\citenamefont{Kells,
  Meidan, and Brouwer}}]{Kells12}
\bibinfo{author}{\bibfnamefont{G.}~\bibnamefont{Kells}},
  \bibinfo{author}{\bibfnamefont{D.}~\bibnamefont{Meidan}}, \bibnamefont{and}
  \bibinfo{author}{\bibfnamefont{P.~W.} \bibnamefont{Brouwer}},
  \bibinfo{journal}{Phys. Rev. B} \textbf{\bibinfo{volume}{86}},
  \bibinfo{pages}{100503} (\bibinfo{year}{2012}{\natexlab{b}}).

\bibitem[{\citenamefont{Liu et~al.}(2017)\citenamefont{Liu, Setiawan, Sau, and
  Das~Sarma}}]{Liu17}
\bibinfo{author}{\bibfnamefont{C.-X.} \bibnamefont{Liu}},
  \bibinfo{author}{\bibfnamefont{F.}~\bibnamefont{Setiawan}},
  \bibinfo{author}{\bibfnamefont{J.~D.} \bibnamefont{Sau}}, \bibnamefont{and}
  \bibinfo{author}{\bibfnamefont{S.}~\bibnamefont{Das~Sarma}},
  \bibinfo{journal}{Phys. Rev. B} \textbf{\bibinfo{volume}{96}},
  \bibinfo{pages}{054520} (\bibinfo{year}{2017}).

\bibitem[{\citenamefont{Benjamin and Pachos}(2010)}]{Benjamin10}
\bibinfo{author}{\bibfnamefont{C.}~\bibnamefont{Benjamin}} \bibnamefont{and}
  \bibinfo{author}{\bibfnamefont{J.~K.} \bibnamefont{Pachos}},
  \bibinfo{journal}{Phys. Rev. B} \textbf{\bibinfo{volume}{81}},
  \bibinfo{pages}{085101} (\bibinfo{year}{2010}).

\bibitem[{\citenamefont{En-Ming et~al.}(2014)\citenamefont{En-Ming, Yi-Ming,
  Lu-Bing, and Bai-Gen}}]{EnMing14}
\bibinfo{author}{\bibfnamefont{S.}~\bibnamefont{En-Ming}},
  \bibinfo{author}{\bibfnamefont{P.}~\bibnamefont{Yi-Ming}},
  \bibinfo{author}{\bibfnamefont{S.}~\bibnamefont{Lu-Bing}}, \bibnamefont{and}
  \bibinfo{author}{\bibfnamefont{W.}~\bibnamefont{Bai-Gen}},
  \bibinfo{journal}{Chinese Physics B} \textbf{\bibinfo{volume}{23}},
  \bibinfo{pages}{057201} (\bibinfo{year}{2014}).

\bibitem[{\citenamefont{Sun and Wu}(2014)}]{Sun14}
\bibinfo{author}{\bibfnamefont{B.~Y.} \bibnamefont{Sun}} \bibnamefont{and}
  \bibinfo{author}{\bibfnamefont{M.~W.} \bibnamefont{Wu}},
  \bibinfo{journal}{New Journal of Physics} \textbf{\bibinfo{volume}{16}},
  \bibinfo{pages}{073045} (\bibinfo{year}{2014}).

\bibitem[{\citenamefont{Ueda and Yokoyama}(2014{\natexlab{a}})}]{Ueda14}
\bibinfo{author}{\bibfnamefont{A.}~\bibnamefont{Ueda}} \bibnamefont{and}
  \bibinfo{author}{\bibfnamefont{T.}~\bibnamefont{Yokoyama}},
  \bibinfo{journal}{Phys. Rev. B} \textbf{\bibinfo{volume}{90}},
  \bibinfo{pages}{081405} (\bibinfo{year}{2014}{\natexlab{a}}).

\bibitem[{\citenamefont{Ueda and Yokoyama}(2014{\natexlab{b}})}]{Ueda14b}
\bibinfo{author}{\bibfnamefont{A.}~\bibnamefont{Ueda}} \bibnamefont{and}
  \bibinfo{author}{\bibfnamefont{T.}~\bibnamefont{Yokoyama}},
  \bibinfo{journal}{Physics Procedia} \textbf{\bibinfo{volume}{58}},
  \bibinfo{pages}{182} (\bibinfo{year}{2014}{\natexlab{b}}).

\bibitem[{\citenamefont{Tripathi et~al.}(2016)\citenamefont{Tripathi, Das, and
  Rao}}]{Tripathi16}
\bibinfo{author}{\bibfnamefont{K.~M.} \bibnamefont{Tripathi}},
  \bibinfo{author}{\bibfnamefont{S.}~\bibnamefont{Das}}, \bibnamefont{and}
  \bibinfo{author}{\bibfnamefont{S.}~\bibnamefont{Rao}},
  \bibinfo{journal}{Phys. Rev. Lett.} \textbf{\bibinfo{volume}{116}},
  \bibinfo{pages}{166401} (\bibinfo{year}{2016}).

\bibitem[{\citenamefont{Dahan et~al.}(2017)\citenamefont{Dahan, Ahari, Ortiz,
  Seradjeh, and Grosfeld}}]{Dahan17}
\bibinfo{author}{\bibfnamefont{D.}~\bibnamefont{Dahan}},
  \bibinfo{author}{\bibfnamefont{M.~T.} \bibnamefont{Ahari}},
  \bibinfo{author}{\bibfnamefont{G.}~\bibnamefont{Ortiz}},
  \bibinfo{author}{\bibfnamefont{B.}~\bibnamefont{Seradjeh}}, \bibnamefont{and}
  \bibinfo{author}{\bibfnamefont{E.}~\bibnamefont{Grosfeld}},
  \bibinfo{journal}{Phys. Rev. B} \textbf{\bibinfo{volume}{95}},
  \bibinfo{pages}{201114} (\bibinfo{year}{2017}).

\bibitem[{\citenamefont{Fu}(2010)}]{Fu10}
\bibinfo{author}{\bibfnamefont{L.}~\bibnamefont{Fu}}, \bibinfo{journal}{Phys.
  Rev. Lett.} \textbf{\bibinfo{volume}{104}}, \bibinfo{pages}{056402}
  (\bibinfo{year}{2010}).

\bibitem[{\citenamefont{Yamakage and Sato}(2014)}]{Yamakage14}
\bibinfo{author}{\bibfnamefont{A.}~\bibnamefont{Yamakage}} \bibnamefont{and}
  \bibinfo{author}{\bibfnamefont{M.}~\bibnamefont{Sato}},
  \bibinfo{journal}{Physica E: Low-dimensional Systems and Nanostructures}
  \textbf{\bibinfo{volume}{55}}, \bibinfo{pages}{13 } (\bibinfo{year}{2014}).

\bibitem[{\citenamefont{Sau et~al.}(2015)\citenamefont{Sau, Swingle, and
  Tewari}}]{Sau15Nonlocal}
\bibinfo{author}{\bibfnamefont{J.~D.} \bibnamefont{Sau}},
  \bibinfo{author}{\bibfnamefont{B.}~\bibnamefont{Swingle}}, \bibnamefont{and}
  \bibinfo{author}{\bibfnamefont{S.}~\bibnamefont{Tewari}},
  \bibinfo{journal}{Phys. Rev. B} \textbf{\bibinfo{volume}{92}},
  \bibinfo{pages}{020511} (\bibinfo{year}{2015}).

\bibitem[{\citenamefont{Rubbert and Akhmerov}(2016)}]{Rubbert16}
\bibinfo{author}{\bibfnamefont{S.}~\bibnamefont{Rubbert}} \bibnamefont{and}
  \bibinfo{author}{\bibfnamefont{A.~R.} \bibnamefont{Akhmerov}},
  \bibinfo{journal}{Phys. Rev. B} \textbf{\bibinfo{volume}{94}},
  \bibinfo{pages}{115430} (\bibinfo{year}{2016}).

\bibitem[{\citenamefont{Chiu et~al.}(2017)\citenamefont{Chiu, Sau, and
  Sarma}}]{Chiu17}
\bibinfo{author}{\bibfnamefont{C.-K.} \bibnamefont{Chiu}},
  \bibinfo{author}{\bibfnamefont{J.~D.} \bibnamefont{Sau}}, \bibnamefont{and}
  \bibinfo{author}{\bibfnamefont{S.~D.} \bibnamefont{Sarma}},
  \bibinfo{journal}{arXiv preprint arXiv:1709.04475}  (\bibinfo{year}{2017}).

\bibitem[{\citenamefont{Plugge et~al.}(2017)\citenamefont{Plugge, Rasmussen,
  Egger, and Flensberg}}]{Plugge17}
\bibinfo{author}{\bibfnamefont{S.}~\bibnamefont{Plugge}},
  \bibinfo{author}{\bibfnamefont{A.}~\bibnamefont{Rasmussen}},
  \bibinfo{author}{\bibfnamefont{A.}~\bibnamefont{Egger}}, \bibnamefont{and}
  \bibinfo{author}{\bibfnamefont{K.}~\bibnamefont{Flensberg}},
  \bibinfo{journal}{New Journal of Physics} \textbf{\bibinfo{volume}{19}},
  \bibinfo{pages}{012001} (\bibinfo{year}{2017}).

\bibitem[{\citenamefont{Plugge et~al.}(2016)\citenamefont{Plugge, Landau, Sela,
  Altland, Flensberg, and Egger}}]{Plugge16}
\bibinfo{author}{\bibfnamefont{S.}~\bibnamefont{Plugge}},
  \bibinfo{author}{\bibfnamefont{L.~A.} \bibnamefont{Landau}},
  \bibinfo{author}{\bibfnamefont{E.}~\bibnamefont{Sela}},
  \bibinfo{author}{\bibfnamefont{A.}~\bibnamefont{Altland}},
  \bibinfo{author}{\bibfnamefont{K.}~\bibnamefont{Flensberg}},
  \bibnamefont{and} \bibinfo{author}{\bibfnamefont{R.}~\bibnamefont{Egger}},
  \bibinfo{journal}{Phys. Rev. B} \textbf{\bibinfo{volume}{94}},
  \bibinfo{pages}{174514} (\bibinfo{year}{2016}).

\bibitem[{\citenamefont{Hell et~al.}(2017{\natexlab{a}})\citenamefont{Hell,
  Flensberg, and Leijnse}}]{Hell17csup}
\bibinfo{author}{\bibfnamefont{M.}~\bibnamefont{Hell}},
  \bibinfo{author}{\bibfnamefont{K.}~\bibnamefont{Flensberg}},
  \bibnamefont{and} \bibinfo{author}{\bibfnamefont{M.}~\bibnamefont{Leijnse}},
  \bibinfo{journal}{Supplemental material to "Distinguishing Majorana bound
  states from localized Andreev bound states by interferometry"}
  (\bibinfo{year}{2017}{\natexlab{a}}).

\bibitem[{\citenamefont{Yang et~al.}(2002)\citenamefont{Yang, Yang, Cheng, and
  Culbertson}}]{Yang02}
\bibinfo{author}{\bibfnamefont{C.~H.} \bibnamefont{Yang}},
  \bibinfo{author}{\bibfnamefont{M.~J.} \bibnamefont{Yang}},
  \bibinfo{author}{\bibfnamefont{K.~A.} \bibnamefont{Cheng}}, \bibnamefont{and}
  \bibinfo{author}{\bibfnamefont{J.~C.} \bibnamefont{Culbertson}},
  \bibinfo{journal}{Phys. Rev. B} \textbf{\bibinfo{volume}{66}},
  \bibinfo{pages}{115306} (\bibinfo{year}{2002}).

\bibitem[{\citenamefont{Ren et~al.}(2013)\citenamefont{Ren, Heremans, Gaspe,
  Vijeyaragunathan, Mishima, and Santos}}]{Ren13}
\bibinfo{author}{\bibfnamefont{S.~L.} \bibnamefont{Ren}},
  \bibinfo{author}{\bibfnamefont{J.~J.} \bibnamefont{Heremans}},
  \bibinfo{author}{\bibfnamefont{C.~K.} \bibnamefont{Gaspe}},
  \bibinfo{author}{\bibfnamefont{S.}~\bibnamefont{Vijeyaragunathan}},
  \bibinfo{author}{\bibfnamefont{T.~D.} \bibnamefont{Mishima}},
  \bibnamefont{and} \bibinfo{author}{\bibfnamefont{M.~B.}
  \bibnamefont{Santos}}, \bibinfo{journal}{Journal of Physics: Condensed
  Matter} \textbf{\bibinfo{volume}{25}}, \bibinfo{pages}{435301}
  (\bibinfo{year}{2013}).

\bibitem[{\citenamefont{Akera}(1993)}]{Akera93}
\bibinfo{author}{\bibfnamefont{H.}~\bibnamefont{Akera}},
  \bibinfo{journal}{Phys. Rev. B} \textbf{\bibinfo{volume}{47}},
  \bibinfo{pages}{6835} (\bibinfo{year}{1993}).

\bibitem[{\citenamefont{Higginbotham et~al.}(2015)\citenamefont{Higginbotham,
  Albrecht, Kirsanskas, Chang, Kuemmeth, Krogstrup, Jespersen, Nygard,
  Flensberg, and Marcus}}]{Higginbotham15}
\bibinfo{author}{\bibfnamefont{A.~P.} \bibnamefont{Higginbotham}},
  \bibinfo{author}{\bibfnamefont{S.~M.} \bibnamefont{Albrecht}},
  \bibinfo{author}{\bibfnamefont{G.}~\bibnamefont{Kirsanskas}},
  \bibinfo{author}{\bibfnamefont{W.}~\bibnamefont{Chang}},
  \bibinfo{author}{\bibfnamefont{F.}~\bibnamefont{Kuemmeth}},
  \bibinfo{author}{\bibfnamefont{P.}~\bibnamefont{Krogstrup}},
  \bibinfo{author}{\bibfnamefont{T.~S.} \bibnamefont{Jespersen}},
  \bibinfo{author}{\bibfnamefont{J.}~\bibnamefont{Nygard}},
  \bibinfo{author}{\bibfnamefont{K.}~\bibnamefont{Flensberg}},
  \bibnamefont{and} \bibinfo{author}{\bibfnamefont{C.~M.}
  \bibnamefont{Marcus}}, \bibinfo{journal}{Nat. Phys.}
  \textbf{\bibinfo{volume}{11}}, \bibinfo{pages}{1017} (\bibinfo{year}{2015}).

\bibitem[{\citenamefont{Bretheau et~al.}(2013)\citenamefont{Bretheau, Girit,
  Urbina, Esteve, and Pothier}}]{Bretheau13}
\bibinfo{author}{\bibfnamefont{L.}~\bibnamefont{Bretheau}},
  \bibinfo{author}{\bibfnamefont{C.~O.} \bibnamefont{Girit}},
  \bibinfo{author}{\bibfnamefont{C.}~\bibnamefont{Urbina}},
  \bibinfo{author}{\bibfnamefont{D.}~\bibnamefont{Esteve}}, \bibnamefont{and}
  \bibinfo{author}{\bibfnamefont{H.}~\bibnamefont{Pothier}},
  \bibinfo{journal}{Phys. Rev. X} \textbf{\bibinfo{volume}{3}},
  \bibinfo{pages}{041034} (\bibinfo{year}{2013}).

\bibitem[{\citenamefont{Patel et~al.}(2016)\citenamefont{Patel, Pechenezhskiy,
  Plourde, Vavilov, and McDermott}}]{Patel16}
\bibinfo{author}{\bibfnamefont{U.}~\bibnamefont{Patel}},
  \bibinfo{author}{\bibfnamefont{I.~V.} \bibnamefont{Pechenezhskiy}},
  \bibinfo{author}{\bibfnamefont{B.}~\bibnamefont{Plourde}},
  \bibinfo{author}{\bibfnamefont{M.}~\bibnamefont{Vavilov}}, \bibnamefont{and}
  \bibinfo{author}{\bibfnamefont{R.}~\bibnamefont{McDermott}},
  \bibinfo{journal}{arXiv preprint arXiv:1610.09351}  (\bibinfo{year}{2016}).

\bibitem[{\citenamefont{Hell et~al.}(2017{\natexlab{b}})\citenamefont{Hell,
  Leijnse, and Flensberg}}]{Hell17a}
\bibinfo{author}{\bibfnamefont{M.}~\bibnamefont{Hell}},
  \bibinfo{author}{\bibfnamefont{M.}~\bibnamefont{Leijnse}}, \bibnamefont{and}
  \bibinfo{author}{\bibfnamefont{K.}~\bibnamefont{Flensberg}},
  \bibinfo{journal}{Phys. Rev. Lett.} \textbf{\bibinfo{volume}{118}},
  \bibinfo{pages}{107701} (\bibinfo{year}{2017}{\natexlab{b}}).

\bibitem[{\citenamefont{Lutchyn et~al.}(2011)\citenamefont{Lutchyn, Stanescu,
  and Das~Sarma}}]{Lutchyn11Multiband}
\bibinfo{author}{\bibfnamefont{R.~M.} \bibnamefont{Lutchyn}},
  \bibinfo{author}{\bibfnamefont{T.~D.} \bibnamefont{Stanescu}},
  \bibnamefont{and}
  \bibinfo{author}{\bibfnamefont{S.}~\bibnamefont{Das~Sarma}},
  \bibinfo{journal}{Phys. Rev. Lett.} \textbf{\bibinfo{volume}{106}},
  \bibinfo{pages}{127001} (\bibinfo{year}{2011}).

\bibitem[{\citenamefont{Potter and Lee}(2010)}]{Multichannel1}
\bibinfo{author}{\bibfnamefont{A.~C.} \bibnamefont{Potter}} \bibnamefont{and}
  \bibinfo{author}{\bibfnamefont{P.~A.} \bibnamefont{Lee}},
  \bibinfo{journal}{Phys. Rev. Lett.} \textbf{\bibinfo{volume}{105}},
  \bibinfo{pages}{227003} (\bibinfo{year}{2010}).

\bibitem[{\citenamefont{Stanescu et~al.}(2011)\citenamefont{Stanescu, Lutchyn,
  and {Das Sarma}}}]{Disorder7}
\bibinfo{author}{\bibfnamefont{T.~D.} \bibnamefont{Stanescu}},
  \bibinfo{author}{\bibfnamefont{R.~M.} \bibnamefont{Lutchyn}},
  \bibnamefont{and} \bibinfo{author}{\bibfnamefont{S.}~\bibnamefont{{Das
  Sarma}}}, \bibinfo{journal}{Phys. Rev. B} \textbf{\bibinfo{volume}{84}},
  \bibinfo{pages}{144522} (\bibinfo{year}{2011}).

\end{thebibliography}


\begin{thebibliography}{4}
\expandafter\ifx\csname natexlab\endcsname\relax\def\natexlab#1{#1}\fi
\expandafter\ifx\csname bibnamefont\endcsname\relax
  \def\bibnamefont#1{#1}\fi
\expandafter\ifx\csname bibfnamefont\endcsname\relax
  \def\bibfnamefont#1{#1}\fi
\expandafter\ifx\csname citenamefont\endcsname\relax
  \def\citenamefont#1{#1}\fi
\expandafter\ifx\csname url\endcsname\relax
  \def\url#1{\texttt{#1}}\fi
\expandafter\ifx\csname urlprefix\endcsname\relax\def\urlprefix{URL }\fi
\providecommand{\bibinfo}[2]{#2}
\providecommand{\eprint}[2][]{\url{#2}}

\bibitem[{\citenamefont{Prada et~al.}(2017)\citenamefont{Prada, Aguado, and
  San-Jose}}]{Prada17}
\bibinfo{author}{\bibfnamefont{E.}~\bibnamefont{Prada}},
  \bibinfo{author}{\bibfnamefont{R.}~\bibnamefont{Aguado}}, \bibnamefont{and}
  \bibinfo{author}{\bibfnamefont{P.}~\bibnamefont{San-Jose}},
  \bibinfo{journal}{Phys. Rev. B} \textbf{\bibinfo{volume}{96}},
  \bibinfo{pages}{085418} (\bibinfo{year}{2017}).

\bibitem[{\citenamefont{Hell et~al.}(2017)\citenamefont{Hell, Flensberg, and
  Leijnse}}]{Hell17b}
\bibinfo{author}{\bibfnamefont{M.}~\bibnamefont{Hell}},
  \bibinfo{author}{\bibfnamefont{K.}~\bibnamefont{Flensberg}},
  \bibnamefont{and} \bibinfo{author}{\bibfnamefont{M.}~\bibnamefont{Leijnse}},
  \bibinfo{journal}{Phys. Rev. B} \textbf{\bibinfo{volume}{96}},
  \bibinfo{pages}{035444} (\bibinfo{year}{2017}).

\bibitem[{\citenamefont{Bruus and Flensberg}(2004)}]{Bruus04}
\bibinfo{author}{\bibfnamefont{H.}~\bibnamefont{Bruus}} \bibnamefont{and}
  \bibinfo{author}{\bibfnamefont{K.}~\bibnamefont{Flensberg}},
  \emph{\bibinfo{title}{Many-body quantum theory in condensed matter physics:
  an introduction}} (\bibinfo{publisher}{Oxford University Press},
  \bibinfo{year}{2004}).

\bibitem[{\citenamefont{Plugge et~al.}(2017)\citenamefont{Plugge, Rasmussen,
  Egger, and Flensberg}}]{Plugge17}
\bibinfo{author}{\bibfnamefont{S.}~\bibnamefont{Plugge}},
  \bibinfo{author}{\bibfnamefont{A.}~\bibnamefont{Rasmussen}},
  \bibinfo{author}{\bibfnamefont{A.}~\bibnamefont{Egger}}, \bibnamefont{and}
  \bibinfo{author}{\bibfnamefont{K.}~\bibnamefont{Flensberg}},
  \bibinfo{journal}{New Journal of Physics} \textbf{\bibinfo{volume}{19}},
  \bibinfo{pages}{012001} (\bibinfo{year}{2017}).

\end{thebibliography}
\end{bibunit}

\end{document}